\documentclass{aastex62}
\shorttitle{Lorentz Factor Evolution of an Expanding Jet Shell}
\shortauthors{Lin et al.}
\begin{document}
\title{Lorentz Factor Evolution of an Expanding Jet Shell Observed in Gamma-ray Burst: Case study of GRB 160625B}
\correspondingauthor{Rui-Jing Lu, Da-Bin Lin}
\email{luruijing@gxu.edu.cn, lindabin@gxu.edu.cn}
\author{Da-Bin Lin}
\affil{Laboratory for Relativistic Astrophysics, Department of Physics, Guangxi University, Nanning 530004, China}
\author{Rui-Jing Lu}
\affil{Laboratory for Relativistic Astrophysics, Department of Physics, Guangxi University, Nanning 530004, China}
\author{Shen-Shi Du}
\affil{Laboratory for Relativistic Astrophysics, Department of Physics, Guangxi University, Nanning 530004, China}
\author{Tong Liu}
\affil{Department of Astronomy, Xiamen University, Xiamen, Fujian 361005, China}
\author{Xiang-Gao Wang}
\affil{Laboratory for Relativistic Astrophysics, Department of Physics, Guangxi University, Nanning 530004, China}
\author{En-Wei Liang}
\affil{Laboratory for Relativistic Astrophysics, Department of Physics, Guangxi University, Nanning 530004, China}

\begin{abstract}
The Lorentz factor of a relativistic jet and its evolution during the jet expansion are difficult to estimate,
especially for the jets in gamma-ray bursts (GRBs).
However, it is related to the understanding of jet physics.
Owing to the absorption of two-photon pair production ($\gamma\gamma{\leftrightarrow}e^+e^-$),
a high-energy spectral cutoff may appear in the radiation spectrum of GRBs.
We search such kind of high-energy cutoff in GRB~160625B, which is one of the brightest bursts in recent years.
It is found that the high-energy spectral cutoff is obvious for the first pulse in the second emission episode of GRB~160625B
(i.e., $\sim186-192$~s after the burst first trigger),
which is smooth and well-shaped.
Then, we estimate the Lorentz factor and radiation location of the jet shell
associated with the first pulse in the second emission episode of GRB~160625B.
It is found that the radiation location increases with time.
In addition, the Lorentz factor remains almost constant during the expansion of the jet shell.
This reveals that the magnetization of the jet is low or intermediate in the emission region,
event though the jet could be still Poynting flux dominated at smaller radii to avoid a bright thermal component in the emission episode.
\end{abstract}
\section{Introduction} \label{sec:intro}
Gamma-ray bursts (GRBs) are the most powerful explosions of $\gamma$-rays in the Universe.
It was early realized that the phenomena of GRBs are associated with an ultrarelativistic jet
(\citealp{Krolik_JH-1991-Pier_EA, Fenimore_EE-1993-Epstein_RI, Woods_E-1995-Loeb_A,Baring_MG-1997-Harding_AK}).
However, the Lorentz factor ($\Gamma$) evolution of an expanding GRB jet was not observed
since the discovery of GRBs in the late 1960s.
A GRB jet can be either matter dominated or Poynting flux dominated.
A matter dominated GRB jet, also called ``fireball'', expands under of its thermal pressure.
The Thomson scattering optical depth decreases during the jet expansion
and the thermal photons are released near the photosphere.
Then, a bright quasi-thermal spectral component,
usually accompanied with a non-thermal component formed in the internal shocks, is expected
(\citealp{Goodman_J-1986,Paczynski_B-1986,Thompson_C-1994,
Meszaros_P-2000-Rees_MJ,Rees_MJ-1994-Meszaros_P,Meszaros_P-2002-Ramirez-Ruiz_E,Toma_K-2011-Wu_XF,Pe'er_A-2012-Zhang_BB}).
In a Poynting flux dominated jet,
the magnetic energy is discharged via magnetic reconnection
(e.g., \citealp{Spruit_HC-2001-Daigne_F,Drenkhahn_G-2002-Spruit_HC,Giannios_D_-2008-A&A.480.305G,
Zhang_B-2011-Yan_H,McKinney_JC-2012-Uzdensky_DA-MNRAS.419.573M,Kumar_P-2015-Crumley_Patrick_-MNRAS.453.1820K,
Sironi_L-2016-Giannios_D-MNRAS.462.48S,Beniamini_P-2016-Granot_J-MNRAS.459.3635B,Granot_J-2016-ApJ.816L.20G}
).
A part of the dissipated magnetic energy is used to accelerate the jet to the high $\Gamma$
and the other accelerates electrons to relativistic energies.
The accelerated electrons gyrate in the magnetic fields and thus
the photons are formed via synchrotron or inverse-Compton radiation processes.
The radiation of a Poynting flux dominated jet is associated with the acceleration of the jet.
This behavior is different from the radiation behavior in the internal shocks,
of which the Lorentz factor remains almost constant.
Thus, a direct observation of the Lorentz factor and its evolution
during the jet expansion can help to
clarify the jet physics in GRBs.

Several methods have been proposed to infer the Lorentz factor of a GRB jet.
The widely used method is based on the onset bump of the afterglows (\citealp{Sari_R-1999b-Piran_T,Liang_EW-2010-Yi_SX,Liang_EW-2015-Lin_TT,Ghirlanda_G-2012-Nava_L}),
of which the peak time is related to the bulk Lorentz factor $\Gamma_0$ of
the jets after producing the prompt $\gamma$-rays.
This value can be somewhat smaller (for internal shocks) or larger
(for Poynting flux dissipation, \citealp{Zhang_B-2014-Zhang_B-ApJ.782.92Z}) than
that measured during the prompt phase.
Moreover, the value of $\Gamma_0$ corresponds to the mean value of jets' Lorentz factor after the prompt emission phase
and could not provide any information about the $\Gamma$ evolution of an expanding jet shell.
The high-energy spectral cutoff induced by the absorption of two-photon pair production  ($\gamma\gamma{\leftrightarrow}e^+e^-$)
is also used to estimate the Lorentz factor of GRB jet
(\citealp{Krolik_JH-1991-Pier_EA, Fenimore_EE-1993-Epstein_RI, Woods_E-1995-Loeb_A,Baring_MG-1997-Harding_AK};
\citealp{Lithwick_Y-2001-Sari_R}; \citealp{Baring_MG-2006-ApJ.650.1004B}; \citealp{Ackermann_M-2011-Ajello_M,Ackermann_M-2013-Ajello_M,Tang_QW-2015-Peng_FK}).
Different from the first method,
the information about the Lorentz factor evolution during the jet expansion can be inferred
in this method.
Thanks to the broadband spectral coverage of
the Gamma-ray Burst Monitor (GBM, \citealp{Meegan_C-2009-Lichti_G})
and the Large Area Telescope (LAT, \citealp{Atwood_WB-2009-Abdo_AA}) instruments onboard the \emph{Fermi} satellite,
the search for such a high-energy spectral cutoff becomes possible (\citealp{Ackermann_M-2011-Ajello_M,Ackermann_M-2013-Ajello_M,Tang_QW-2015-Peng_FK}).
We search such kind of high-energy spectral cutoff in GRB 160625B, which is one of the brightest bursts in recent years.
It is found that the high-energy spectral cutoff is obvious for
the first pulse in the second emission episode of GRB~160625B
(i.e., $\sim186-192$s after the burst first trigger),
which is very smooth and well-shaped (see Figure~\ref{Fig:LC}).
Then, we estimate the Lorentz factor and radiation location of the jet shell
associated with the first pulse in the second emission episode (FP2EE) of GRB~160625B.

The paper is organized as follows.
The data reduction and joint spectral fittings are performed in Section~\ref{Sec:Data},
where we focus our attention on the FP2EE of GRB~160625B.
According to the obtained results from joint spectral fittings,
the Lorentz factor and radiation location of the radiating jet shell
associated with the FP2EE are estimated in Section~\ref{Sec:Results}.
Here, we assume that the two-photon pair production is responsible for
the formation of the high-energy spectral cutoff.
In Section~\ref{Sec:Conclusion}, we summarize our conclusion.

\section{Data Reduction and Joint Spectral Fittings}\label{Sec:Data}
\begin{figure}
\centering
\begin{tabular}{c}
\includegraphics[scale=0.35, trim=0 0 0 0, clip]{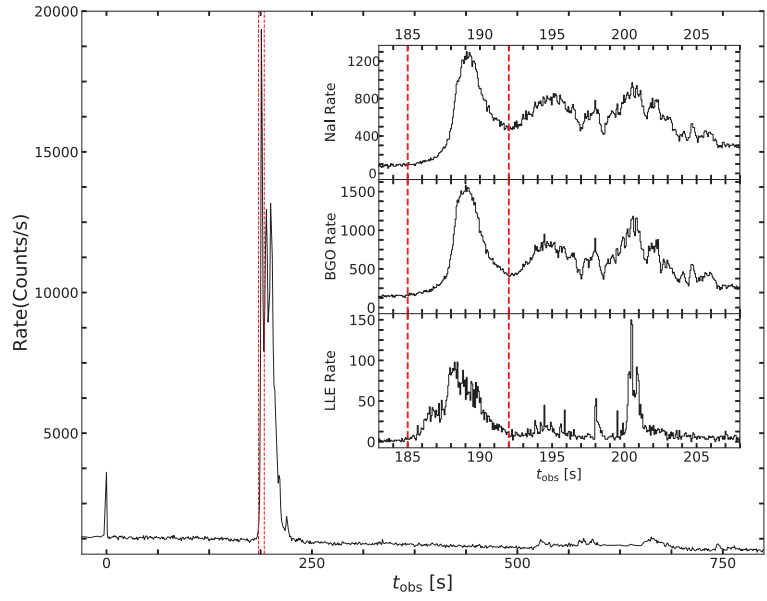}
\end{tabular}
\caption{Light curves of GRB~160625B,
where the two vertical dashed lines mark the time period for our analysis
and the inset shows a zoom around our interested time period.}\label{Fig:LC}
\end{figure}

In our spectral analysis, we use the data from both the GBM and LAT instruments.
GBM has 12 sodium iodide (NaI) scintillation detectors covering the 8~keV-1~MeV energy band,
and two bismuth germanate (BGO) scintillation detectors being sensitive to the 200keV-40MeV energy band
(\citealp{Meegan_C-2009-Lichti_G}).
The brightest NaI and BGO detectors are used for our analyses.
For the data from LAT instruments,
LAT Low Energy data (LLE \citealp{Pelassa_V-2010-Preece_R,Ajello_M-2014-Albert_A}) are used in our spectral analysis.
The LLE data are obtained by adopting LLE technique (\citealp{Pelassa_V-2010-Preece_R,Ajello_M-2014-Albert_A}),
which is designed to study bright transients in the $\sim30$~MeV-1~GeV energy range
and was successfully applied to GRBs (\citealp{Tang_QW-2015-Peng_FK,Guiriec_S-2015-Kouveliotou_C,Moretti_E-2016-Axelsson_M,Burgess_JM-2016-Begue_D})
and solar flares (\citealp{Ackermann_M-2012-Ajello_M,Ajello_M-2014-Albert_A}) in spectral analysis.
The python source package {\tt gtBurst}\footnote{\url{https://github.com/giacomov/gtburst}}
is used to extract the light curves and source spectra of GBM and LLE from their TTE data, respectively.
Our obtained light curve of the second emission episode in GRB~160625B is shown in Figure~\ref{Fig:LC},
where $t_{\rm obs}$ is the observer time by setting $t_{\rm obs}=0$ at the burst first trigger
(i.e., 22:40:16.28~UT on 25 June 2016 \citealp{Burn_E-2016}).
For the discussions about this burst,
one can refer to, e.g., \cite{Troja_E-2017-Lipunov_VM-Natur.547.425T},
\cite{Lu_HJ-2017-Lu_J}, \cite{Fraija_N-2017-Veres_P}
\cite{Wang_YZ-2017-Wang_H-ApJ.836.81W},
\cite{Alexander_KD-2017-Laskar_T-ApJ.848.69A}, \cite{Zhang_BB-2016-Zhang_B}, and
\cite{Ravasio_ME-2018-Oganesyan_G-A&A.613A.16R}.
From Figure~\ref{Fig:LC}, one can find that the FP2EE (i.e., $\sim186-192$~s,
marked with the two vertical dashed lines)
of GRB~160625B is very smooth and well-shaped,
which is very different from the light curve formed in the photosphere, e.g., GRB~090902B (\citealp{Abdo_AA-2009-Ackermann_M-ApJ.706L.138A}).
Then, we would like to believe that
the FP2EE of GRB~160625B is formed in an expanding jet shell.
The facts used to support this idea will be summarized in the end of Section~\ref{Sec:Results}.

The high-energy spectral cutoff induced by the absorption of two-photon pair production
can be used to estimate the Lorentz factor and radiation location of a GRB jet.
Then, we search such kind of high-energy spectral cutoff in the FP2EE of GRB~160625B.
{\tt XSPEC} (\citealp{Arnaud_KA_-1996-ASPC.101.17A}) is used to perform joint spectral fitting for the data from GBM and LAT instruments,
where {\tt pgstat} is adopted to judge the goodness of the spectral fittings.
In our spectral analysis, we adopt the Band+cutoff spectral model\footnote{\url{https://asd.gsfc.nasa.gov/Takanori.Sakamoto/personal/}}, i.e.,
\begin{equation}\label{Eq:BandCutoff}
{N_E} = N_0\left\{ {\begin{array}{*{20}{c}}
{\left( \frac{E}{1\rm keV}\right)^\alpha\exp \left( { - \frac{E}{{{E_0}}}} \right),}&{E < \frac{{{E_0}{E_{\rm c}}}}{{{E_{\rm c}} - {E_0}}}\left( {\alpha  - \beta } \right),}\\
{{K_2}{\left( \frac{E}{1\rm keV}\right)^\beta }\exp \left( { - \frac{E}{E_{\rm c}}} \right),}&{E > \frac{{{E_0}{E_{\rm c}}}}{{{E_{\rm c}} - {E_0}}}\left( {\alpha  - \beta } \right),}
\end{array}} \right.
\end{equation}
with
\[{K_2} = {\left[ {\frac{{{E_0}{E_{\rm c}}}}{{{E_{\rm c}} - {E_0}}}\left( {\alpha  - \beta } \right)} \right]^{\alpha  - \beta }}\exp \left( {\beta  - \alpha } \right).\]
The time intervals ($\in[186, 192]$~s)
for our spectral analysis can be found in Table~\ref{Tab:Fitting_Results}.
The joint spectral fittings are shown in Figure~\ref{Fig:Joint_Spectral_fittings_LLE} and the obtained results are reported in Table~\ref{Tab:Fitting_Results}.
One can find that the Band+cutoff model well describes the observational data.
In addition, the high-energy spectral cutoff is obvious in each time interval.
We would like to point out that it is safe to use the LLE data
in the spectral analysis for the FP2EE in GRB~160625B.
The reasons are shown in Appendix~\ref{Appendix:A}.

\section{Estimation of $\Gamma$ and $R$}\label{Sec:Results}
In the scenario that the two-photon pair production is responsible for
the formation of the high-energy spectral cutoff,
one can estimate the value of
$\Lambda(t_{\rm obs})\equiv{R(t_{\rm obs})}/[2\Gamma(t_{\rm obs})]^{2\beta}$
(see Appendix~\ref{Appendix:B} for details), i.e.,
\begin{equation}\label{Eq:R_Gamma}
\Lambda  \equiv
\frac{R}{{{{\left( {2\Gamma } \right)}^{2\beta }}}}
=\frac{{{N_0}{K_2}{\sigma _{\rm{T}}}d_L^2}}{2c}{E_{{\rm{ch}}}}(1 - \beta ){(1 + z)^{ - 2\beta  - 3}}{\left( {\frac{{{E_{{\rm{ch}}}}{E_c}}}{{m_e^2{c^4}}}} \right)^{ - 1 - \beta }}F(\beta ),
\end{equation}
where $R(t_{\rm obs})$ is the radiation location of the jet shell (relative to the jet base) at $t_{\rm obs}$,
$\Gamma(t_{\rm obs})$ is the Lorentz factor of the jet shell at $t_{\rm obs}$,
$z=1.406$ (\citealp{Xu_D-2016-Malesani_D}) and $d_L=3.11\times 10^{28}$~cm are the redshift and the luminosity distance of GRB~160625B,
and $\sigma _{\rm{T}}$, $m_{\rm e}$, and $c$ denote  fundamental physical constants with conventional meanings.
The $F( \beta)$ is a function of $\beta$ (\citealp{Abdo_AA-2009-Ackermann_M-Sci.323.1688A})
and can be described as $F( \beta)\approx 0.597(-\beta)^{-2.30}$
for $-2.90\leqslant\beta\leqslant -1.0$ (\citealp{Ackermann_M-2011-Ajello_M}).
For the details of $F(\beta)$, one can refer
the Supporting Online Material of \cite{Abdo_AA-2009-Ackermann_M-Sci.323.1688A}.
The value of $\Lambda(t_{\rm obs})$ is reported in Table~\ref{Tab:Fitting_Results}
and shown in Figure~\ref{Fig:Lambda}.
From this figure,
one can find that the value of $\Lambda$ increases
by four orders of magnitude for the observer time from 187~s to 191~s.
This behavior is the nature outcome of an expanding jet shell
and difficult to be realized for the photospheric emission.
Then, we assume that the FP2EE of GRB~160625B is formed in an expanding jet shell.
The facts used to support this assumption will be summarized in the end of this section.

During the shell's expansion over $\delta t_{\rm obs}$, the jet shell moves from $R({t_{{\rm{obs}}}})$
to $R({t_{{\rm{obs}}}} + \delta t_{\rm obs})$.
In this process, one can have the relation: $dt_{\rm obs}=dR/2\Gamma^2c$
\footnote{As suggested by the referee,
the factor of $1/2$ in the right side of $dt_{\rm obs}=dR/(2\Gamma^2c)$ can be removed by considering
the emission of the entire fluid rather than a fluid element along the line of sight.
Then, we also investigate the situation with $R({t_{{\rm{obs}}}} + \delta t_{\rm obs}) - R({t_{{\rm{obs}}}})
\backsimeq {0.5\times [ {\Gamma ( {{t_{{\rm{obs}}}}})^2 + \Gamma ( {{t_{{\rm{obs}}}} + \delta t_{\rm obs}} )^2} ]}c\delta t_{\rm obs}$.
The obtained result is consistent with a coasting jet shell
and thus does not affect our conclusion in this paper.
To clarify, we take the form of $dt_{\rm obs}=dR/(2\Gamma^2c)$ rather than $dt_{\rm obs}=dR/(\Gamma^2c)$ in this paper.
}, or,
\begin{equation}\label{Eq:Differential_Relation}
R({t_{{\rm{obs}}}} + \delta t_{\rm obs}) - R({t_{{\rm{obs}}}})
\backsimeq {\left[ {\Gamma \left( {{t_{{\rm{obs}}}}} \right)^2 + \Gamma \left( {{t_{{\rm{obs}}}} + \delta t_{\rm obs}} \right)^2} \right]}c\delta t_{\rm obs}.
\end{equation}
With Equation~(\ref{Eq:Differential_Relation}) and a given $\Gamma_{\rm try}=\Gamma(t_{\rm obs}=186.83\rm s)$,
one can calculate the value of $R(t_{\rm obs})$ and $\Gamma(t_{\rm obs})$ at $t_{\rm obs}\neq 186.83\rm s$
by utilizing the value of $\Lambda(t_{\rm obs})$ and $\beta(t_{\rm obs})$ at different $t_{\rm obs}$.
Since the value of $\Gamma_{\rm try}$ could not be obtained previously,
we take $\Gamma_{\rm try}=25$, 50, 100, 125, 250, and 500 for our discussion.
The obtained $\Gamma(t_{\rm obs})$ and $R(t_{\rm obs})$ at different $t_{\rm obs}(\neq 186.83\rm s)$ can be found in Figure~\ref{Fig:R_Gamma},
where the black ``{\small $\blacksquare$}'', red ``$\bullet$'', blue ``$\blacktriangle$'', black ``{\small $\square$}'', red ``$\circ$'', and blue ``$\vartriangle$'' symbols represent the situations
with $\Gamma_{\rm try}=$25, 50, 100, 125, 250, and 500, respectively.
Figure~\ref{Fig:R_Gamma} suggests that the value of $R$ is proportional to $t_{\rm obs}$.
Then, we perform a linear fit on the $R-t_{\rm obs}$ relation for different $\Gamma_{\rm try}$.
The lines of best fit are shown in the right panel of Figure~\ref{Fig:R_Gamma}
and the fitting results are presented in the caption of this figure.
One can find that the value of $R$ is linearly related to $t_{\rm obs}$.
This behavior is consistent with the scenario of an expanding jet shell.
The value of $\Gamma(t_{\rm obs})$ remains almost constant during the expansion of the jet shell
and this behavior does not present significant dependence on the value of $\Gamma_{\rm try}$.
Then, we conclude that the Lorentz factor of the jet shell associated with
the FP2EE is not changed during its expansion.
If the acceleration/deceleration of the jet shell can be described as
$\Gamma  = {\Gamma _0}{\left( {R/{R_0}} \right)^s}$,
one can find the best fitting result of $\Gamma _0$, $R_0$, and $s$
by minimizing the value of $\chi^2$, where
\begin{equation}\label{Eq:chi}
\chi^2=\sum\limits_{t_{\rm obs}=186.83{\rm s}}^{t_{\rm obs}=190.52{\rm s}}\frac{[\log(\Lambda_{\rm mod})-\log(\Lambda)]^2}{\Lambda_{\rm err,log}^2},
\end{equation}
$\Lambda_{\rm err, log}$ is the statistical error of $\log(\Lambda)$,
i.e., $\Lambda_{\rm err, log}=\Lambda_{\rm err}/(\Lambda\ln10)$ with $\Lambda_{\rm err}$ being the statistical error of $\Lambda$,
and $\Lambda_{\rm mod}=R_{\rm mod}/(2\Gamma_{\rm mod})^{2\beta}$ is the model value
calculated based on the value of $\Gamma _0$, $R_0$, and $s$.
The $\Lambda_{\rm mod}(t_{\rm obs})$ is estimated as follows.
Based on the relation of $\Gamma  = {\Gamma _0}{\left( {R/{R_0}} \right)^s}$,
one can have the relation of (\citealp{Lin_DB-2017-Mu_HJ-ApJ.840.95L})
\begin{equation}
t_{\rm obs}-t_0=
\int_{{R_0}}^{R_{\rm mod}} {{{dr} \over {2{\Gamma ^2}c}}}
= \left\{ {\matrix{
   {\left( t_{\rm cm}- t_{\rm c0}\right)/(1 - 2s),\;} & {s \ne 1/2,}  \cr
   {t_{\rm c0}\ln \left( {R_{\rm mod}/{R_0}} \right),} & {s = 1/2,}  \cr
 } } \right.
\end{equation}
or,
\begin{equation}\label{Eq:R_t_obs}
R_{\rm mod}= \left\{ {\matrix{
   {{R_0}{{\left[ {1 + {{t_{\rm obs} - {t_0}} \over {{t_{\rm c0}}}}\left( {1 - 2s} \right)} \right]}^{1/\left( {1 - 2s} \right)}},} & {s \ne 1/2,}  \cr
   {{R_0}\exp \left( {{{t_{\rm obs} - {t_0}} \over {{t_{\rm c0}}}}} \right)} & {s = 1/2,}  \cr
 } } \right.
\end{equation}
where $t_0=186.83$~s, $t_{\rm c0}={R_0}/(2\Gamma _0^2c)$, $t_{\rm cm}=R_{\rm mod}/(2\Gamma_{\rm mod}^2c)$,
and $\Gamma_{\rm mod}  = {\Gamma _0}{\left( {R_{\rm mod}/{R_0}} \right)^s}$.
With the relation of $R_{\rm mod}$ and $t_{\rm obs}$, i.e., Equation~(\ref{Eq:R_t_obs}),
one can easily find the value of $\Lambda_{\rm mod}$ at different observer time $t_{\rm obs}$.
The python source package {\tt SciPy}\footnote{\url{https://github.com/scipy/scipy}}
is used to minimize the $\chi^2$ in Equation~(\ref{Eq:chi}) by implementing the downhill simplex algorithm.
The best fitting result is shown in Figure~\ref{Fig:Lambda} with red line
and read as $\Gamma_0=58$, $R_0=8.27\times 10^{16}$~cm, and $s=9.08\times 10^{-4}$
with $\chi^2=8.42$.
This result is consistent with a coasting jet.

As the end of this section, we now summary the reasons for the assumption
that the FP2EE of GRB~160625B is formed in an expanding jet shell
rather than a streaming outflow:
(I) The FP2EE is very smooth and well-shaped (see Figure~\ref{Fig:LC}),
which is very different from the light curves formed in a photosphere,
e.g., GRB~090902B (\citealp{Abdo_AA-2009-Ackermann_M-ApJ.706L.138A}).
However, smoothness only should not be regarded as a definite criterion.
(II) The value of $\Lambda(t_{\rm obs})$ increases with time
even in the situation that the observed flux decreases significantly.
In addition, the value of $\Lambda(t_{\rm obs})$ increases by $\gtrsim$ four orders of magnitude
from the beginning to the end of the FP2EE.
These behaviors are very difficult to be realized for the radiation from a streaming outflow.

\section{Conclusion}\label{Sec:Conclusion}

In short, we perform a direct estimation of the Lorentz factor and its evolution
for an expanding jet shell in GRBs.
We find that the Lorentz factor of the jet shell associated with
the FP2EE in GRB~160625B remains almost constant during the jet expansion.
This implies that the magnetization of the jet shell associated with the FP2EE
is low or intermediate in the emission region.
With the greatly increased in the spectral coverage,
our method used to estimate the evolution of Lorentz factor for an expanding jet shell
would promote the understanding of the jet dynamics in GRBs.

The magnetic fields in a jet can be dissipated or amplified as the jet expands.
Then, the magnetization of a jet would be a function of the radius.
What we have derived is that the magnetization parameter is low or intermediate in the emission region
rather than in the jet launching region.
It is still possible that the jet being responsible for the FP2EE
may be initially Poynting flux dominated at smaller radii.
With certain initial parameters at the central engine,
it is possible that the jet is initially Poynting flux dominated
but then matter dominated in the emission region
(e.g., \citealp{Gao_H-2015-Zhang_B-ApJ.801.103G}).
The thermal emission from the jet launched with such kind of
initial parameters would be significantly suppressed.
Our spectral fittings reveal that
the spectrum in each time interval of the FP2EE is well described with a Band+cutoff spectral model
rather than a quasi-thermal spectral model (e.g., \citealp{Abdo_AA-2009-Ackermann_M-ApJ.706L.138A})
or a mixture of thermal and non-thermal emission model (e.g., \citealp{Ryde_F-2005-ApJ.625L.95R,Guiriec_S-2011-Connaughton_V-ApJ.727L.33G,Axelsson_M-2012-Baldini_L-ApJ.757L.31A, Guiriec_S-2013-Daigne_F-ApJ.770.32G,Arimoto_M-2016-Asano_K-ApJ.833.139A}).
Then, the jet associated with the FP2EE may be still Poynting flux dominated at smaller radii
in order to avoid a bright thermal component in the FP2EE.

\acknowledgments
We thank the anonymous referee of this work for beneficial
comments that improved the paper.
We also thank Zhang Bing, Gupta Nayantara, and Wu Xue-Feng for helpful discussions.
This work is supported by the National Natural Science Foundation of China
(grant Nos. 11773007, 11533003, 11822304, 11673006),
the Guangxi Science Foundation (grant Nos. 2018GXNSFFA281010, 2016GXNSFDA380027, 2018GXNSFDA281033, 2017AD22006, 2016GXNSFFA380006),
the One-Hundred-Talents Program of Guangxi colleges,
and
High level innovation team and outstanding scholar program in Guangxi colleges.

\software{XSPEC (\citealp{Arnaud_KA_-1996-ASPC.101.17A}), gtBurst ({https://github.com/giacomov/gtburst}), SciPy (https://github.com/scipy/scipy)}
\clearpage
\clearpage
\begin{table}
{\centering
\caption{Fitting results about the FP2EE of GRB~160625B and the obtained $\Lambda$}\label{Tab:Fitting_Results}
\begin{tabular}{c|cccccccc}
\hline\hline
$t_{\rm obs}$	&	Time Interval			&	$\alpha$			&	$\beta$			&	$E_0  ({\rm keV})$			&	$E_1 ({\rm MeV})$			&	$N_0$\tablenotemark{a}			&		$\Lambda ({\rm cm})$					&	$\chi_r^{2}$	\\
\hline																																	186.83 	&$[	186.00 	,	187.50 	]$&$	-0.86 	\pm	0.06 	$&$	-1.56 	\pm	0.09 	$&$	1225 	\pm	393 	$&$	13.9 	\pm	2.0 	$&$	1.86 	\pm	0.49 	$&$	(	1.46 	\pm	1.03 	)\times	10^{23}	$&$	1.01 	$\\
187.84 	&$[	187.20 	,	188.25 	]$&$	-0.69 	\pm	0.03 	$&$	-1.83 	\pm	0.05 	$&$	873 	\pm	98 	$&$	20.3 	\pm	2.2 	$&$	2.78 	\pm	0.41 	$&$	(	2.39 	\pm	1.03 	)\times	10^{24}	$&$	1.09 	$\\
188.08 	&$[	187.71 	,	188.35 	]$&$	-0.69 	\pm	0.03 	$&$	-1.97 	\pm	0.06 	$&$	993 	\pm	96 	$&$	25.4 	\pm	3.4 	$&$	3.99 	\pm	0.54 	$&$	(	1.06 	\pm	0.49 	)\times	10^{25}	$&$	1.06 	$\\
188.27 	&$[	188.12 	,	188.42 	]$&$	-0.68 	\pm	0.03 	$&$	-2.07 	\pm	0.08 	$&$	1033 	\pm	103 	$&$	28.4 	\pm	5.5 	$&$	5.95 	\pm	0.88 	$&$	(	3.63 	\pm	2.18 	)\times	10^{25}	$&$	0.92 	$\\
188.37 	&$[	188.22 	,	188.52 	]$&$	-0.71 	\pm	0.03 	$&$	-2.06 	\pm	0.08 	$&$	1073 	\pm	99 	$&$	25.8 	\pm	4.7 	$&$	8.32 	\pm	1.07 	$&$	(	3.56 	\pm	2.04 	)\times	10^{25}	$&$	1.03 	$\\
188.47 	&$[	188.32 	,	188.62 	]$&$	-0.66 	\pm	0.03 	$&$	-1.99 	\pm	0.07 	$&$	890 	\pm	80 	$&$	22.9 	\pm	3.6 	$&$	7.78 	\pm	1.01 	$&$	(	2.20 	\pm	1.10 	)\times	10^{25}	$&$	1.08 	$\\
188.58 	&$[	188.42 	,	188.72 	]$&$	-0.69 	\pm	0.03 	$&$	-2.05 	\pm	0.07 	$&$	984 	\pm	83 	$&$	23.3 	\pm	3.9 	$&$	9.47 	\pm	1.13 	$&$	(	3.45 	\pm	1.83 	)\times	10^{25}	$&$	1.08 	$\\
188.68 	&$[	188.52 	,	188.82 	]$&$	-0.66 	\pm	0.03 	$&$	-2.01 	\pm	0.07 	$&$	842 	\pm	71 	$&$	19.8 	\pm	3.1 	$&$	9.01 	\pm	1.11 	$&$	(	2.28 	\pm	1.13 	)\times	10^{25}	$&$	0.99 	$\\
188.78 	&$[	188.62 	,	188.92 	]$&$	-0.69 	\pm	0.03 	$&$	-2.13 	\pm	0.07 	$&$	866 	\pm	66 	$&$	29.0 	\pm	5.3 	$&$	10.7 	\pm	1.2 	$&$	(	6.92 	\pm	3.52 	)\times	10^{25}	$&$	0.95 	$\\
188.88 	&$[	188.72 	,	189.02 	]$&$	-0.67 	\pm	0.03 	$&$	-2.11 	\pm	0.06 	$&$	776 	\pm	59 	$&$	27.1 	\pm	4.5 	$&$	10.7 	\pm	1.2 	$&$	(	5.41 	\pm	2.56 	)\times	10^{25}	$&$	1.07 	$\\
188.96 	&$[	188.82 	,	189.12 	]$&$	-0.67 	\pm	0.03 	$&$	-2.09 	\pm	0.06 	$&$	750 	\pm	57 	$&$	25.7 	\pm	4.1 	$&$	11.3 	\pm	1.3 	$&$	(	4.49 	\pm	2.06 	)\times	10^{25}	$&$	0.98 	$\\
189.07 	&$[	188.92 	,	189.22 	]$&$	-0.66 	\pm	0.03 	$&$	-2.14 	\pm	0.06 	$&$	730 	\pm	53 	$&$	26.3 	\pm	4.5 	$&$	10.9 	\pm	1.2 	$&$	(	6.25 	\pm	3.00 	)\times	10^{25}	$&$	1.06 	$\\
189.19 	&$[	189.02 	,	189.32 	]$&$	-0.64 	\pm	0.03 	$&$	-2.23 	\pm	0.06 	$&$	676 	\pm	46 	$&$	34.5 	\pm	7.3 	$&$	10.2 	\pm	1.2 	$&$	(	1.38 	\pm	0.72 	)\times	10^{26}	$&$	1.05 	$\\
189.28 	&$[	189.12 	,	189.42 	]$&$	-0.67 	\pm	0.03 	$&$	-2.33 	\pm	0.07 	$&$	736 	\pm	48 	$&$	41.6 	\pm	10.3 	$&$	11.8 	\pm	1.3 	$&$	(	3.45 	\pm	2.01 	)\times	10^{26}	$&$	1.09 	$\\
189.37 	&$[	189.22 	,	189.52 	]$&$	-0.71 	\pm	0.02 	$&$	-2.34 	\pm	0.07 	$&$	793 	\pm	53 	$&$	38.3 	\pm	9.7 	$&$	13.6 	\pm	1.4 	$&$	(	3.40 	\pm	2.10 	)\times	10^{26}	$&$	1.17 	$\\
189.47 	&$[	189.32 	,	189.62 	]$&$	-0.71 	\pm	0.03 	$&$	-2.29 	\pm	0.08 	$&$	762 	\pm	53 	$&$	26.9 	\pm	5.7 	$&$	13.3 	\pm	1.5 	$&$	(	1.40 	\pm	0.82 	)\times	10^{26}	$&$	1.13 	$\\
189.56 	&$[	189.42 	,	189.72 	]$&$	-0.68 	\pm	0.03 	$&$	-2.22 	\pm	0.07 	$&$	661 	\pm	49 	$&$	24.6 	\pm	4.7 	$&$	12.2 	\pm	1.4 	$&$	(	7.33 	\pm	3.88 	)\times	10^{25}	$&$	1.02 	$\\
189.68 	&$[	189.52 	,	189.82 	]$&$	-0.68 	\pm	0.03 	$&$	-2.24 	\pm	0.07 	$&$	614 	\pm	46 	$&$	30.3 	\pm	6.1 	$&$	11.7 	\pm	1.4 	$&$	(	9.81 	\pm	5.13 	)\times	10^{25}	$&$	1.02 	$\\
189.78 	&$[	189.62 	,	189.92 	]$&$	-0.73 	\pm	0.03 	$&$	-2.24 	\pm	0.07 	$&$	674 	\pm	53 	$&$	33.4 	\pm	7.0 	$&$	13.7 	\pm	1.6 	$&$	(	1.01 	\pm	0.54 	)\times	10^{26}	$&$	1.10 	$\\
189.88 	&$[	189.72 	,	190.02 	]$&$	-0.74 	\pm	0.03 	$&$	-2.32 	\pm	0.07 	$&$	713 	\pm	57 	$&$	50.4 	\pm	13.3 	$&$	13.5 	\pm	1.6 	$&$	(	2.63 	\pm	1.59 	)\times	10^{26}	$&$	1.09 	$\\
189.97 	&$[	189.82 	,	190.12 	]$&$	-0.75 	\pm	0.03 	$&$	-2.30 	\pm	0.07 	$&$	691 	\pm	58 	$&$	44.2 	\pm	12.0 	$&$	12.9 	\pm	1.6 	$&$	(	1.81 	\pm	1.14 	)\times	10^{26}	$&$	1.09 	$\\
190.06 	&$[	189.92 	,	190.22 	]$&$	-0.72 	\pm	0.03 	$&$	-2.42 	\pm	0.06 	$&$	649 	\pm	52 	$&$	89.2 	\pm	28.5 	$&$	10.4 	\pm	1.4 	$&$	(	9.47 	\pm	6.31 	)\times	10^{26}	$&$	1.16 	$\\
190.29 	&$[	190.12 	,	190.48 	]$&$	-0.78 	\pm	0.03 	$&$	-2.51 	\pm	0.05 	$&$	686 	\pm	56 	$&$	139 	\pm	55 	$&$	12.6 	\pm	1.5 	$&$	(	2.69 	\pm	2.10 	)\times	10^{27}	$&$	1.06 	$\\
190.52 	&$[	190.32 	,	190.74 	]$&$	-0.77 	\pm	0.03 	$&$	-2.52 	\pm	0.05 	$&$	565 	\pm	45 	$&$	265 	\pm	161 	$&$	11.0 	\pm	1.4 	$&$	(	5.47 	\pm	5.69 	)\times	10^{27}	$&$	0.97 	$\\
\hline
\end{tabular}}
\tablenotetext{a}{$N_0$ is in the unit of ${\rm photons \cdot cm^{-2}\cdot s^{-1}\cdot keV^{-1}}$}
\end{table}


\clearpage

\clearpage
\begin{figure*}
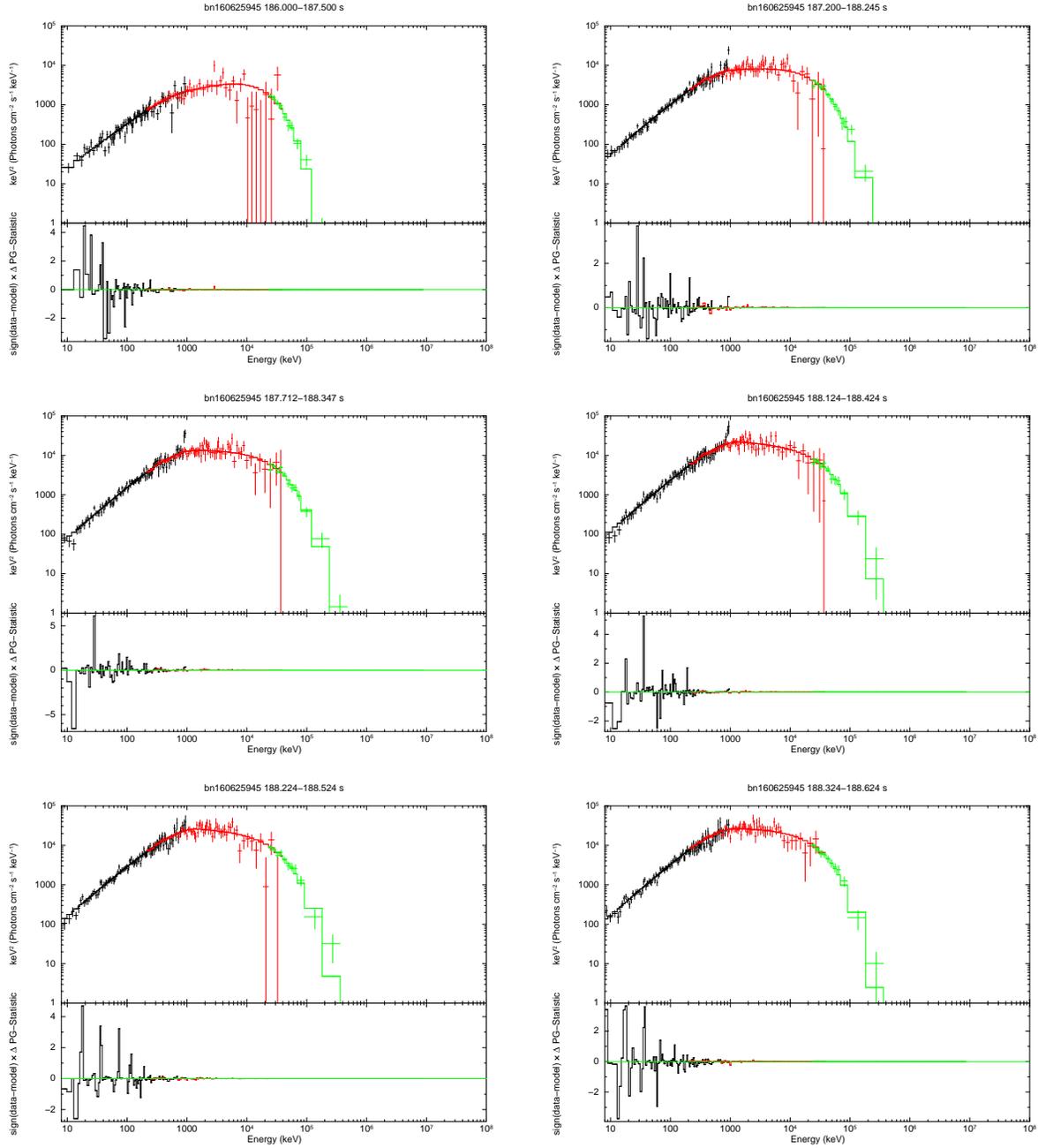

\begin{tabular}{cc}
\includegraphics[angle=270,scale=0.30]{Fig_1_01.eps} &
\includegraphics[angle=270,scale=0.30]{Fig_1_02.eps} \\
\includegraphics[angle=270,scale=0.30]{Fig_1_03.eps} &
\includegraphics[angle=270,scale=0.30]{Fig_1_04.eps} \\
\includegraphics[angle=270,scale=0.30]{Fig_1_05.eps} &
\includegraphics[angle=270,scale=0.30]{Fig_1_06.eps} \\
\end{tabular}
\caption{Spectrum fitting in different time interval for the FP2EE of GRB~160625B,
where the Band+cutoff spectral model is adopted in our spectral fitting.
The data of NaI, BGO, and LLE are shown with black, red, and green ``$+$'' symbols, respectively.
The complete figure set (25 images) is available in the online journal.}
\label{Fig:Joint_Spectral_fittings_LLE}
\end{figure*}
\addtocounter{figure}{-1}
\clearpage
\begin{figure*}
\begin{tabular}{cc}
\includegraphics[angle=270,scale=0.30]{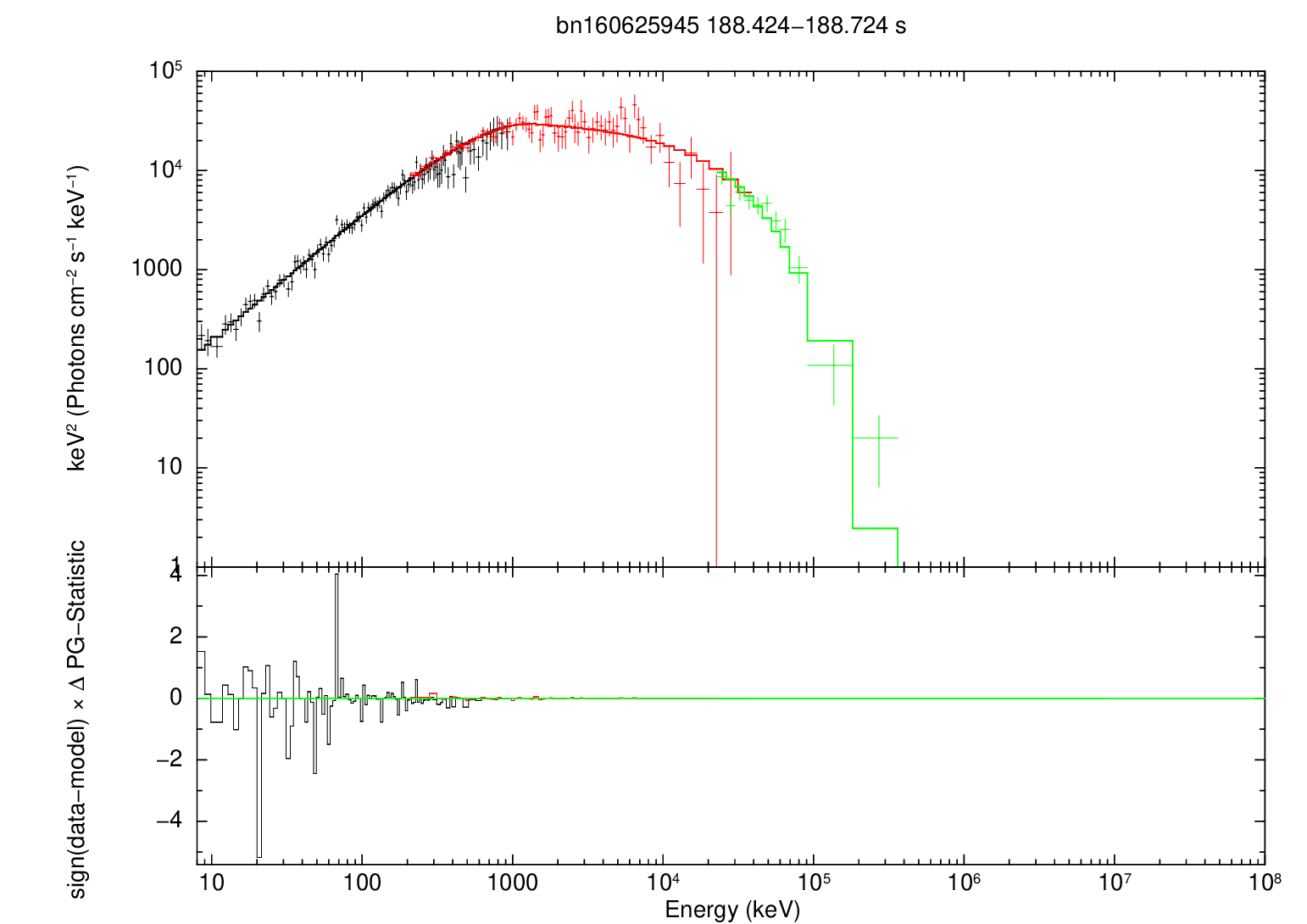} &
\includegraphics[angle=270,scale=0.30]{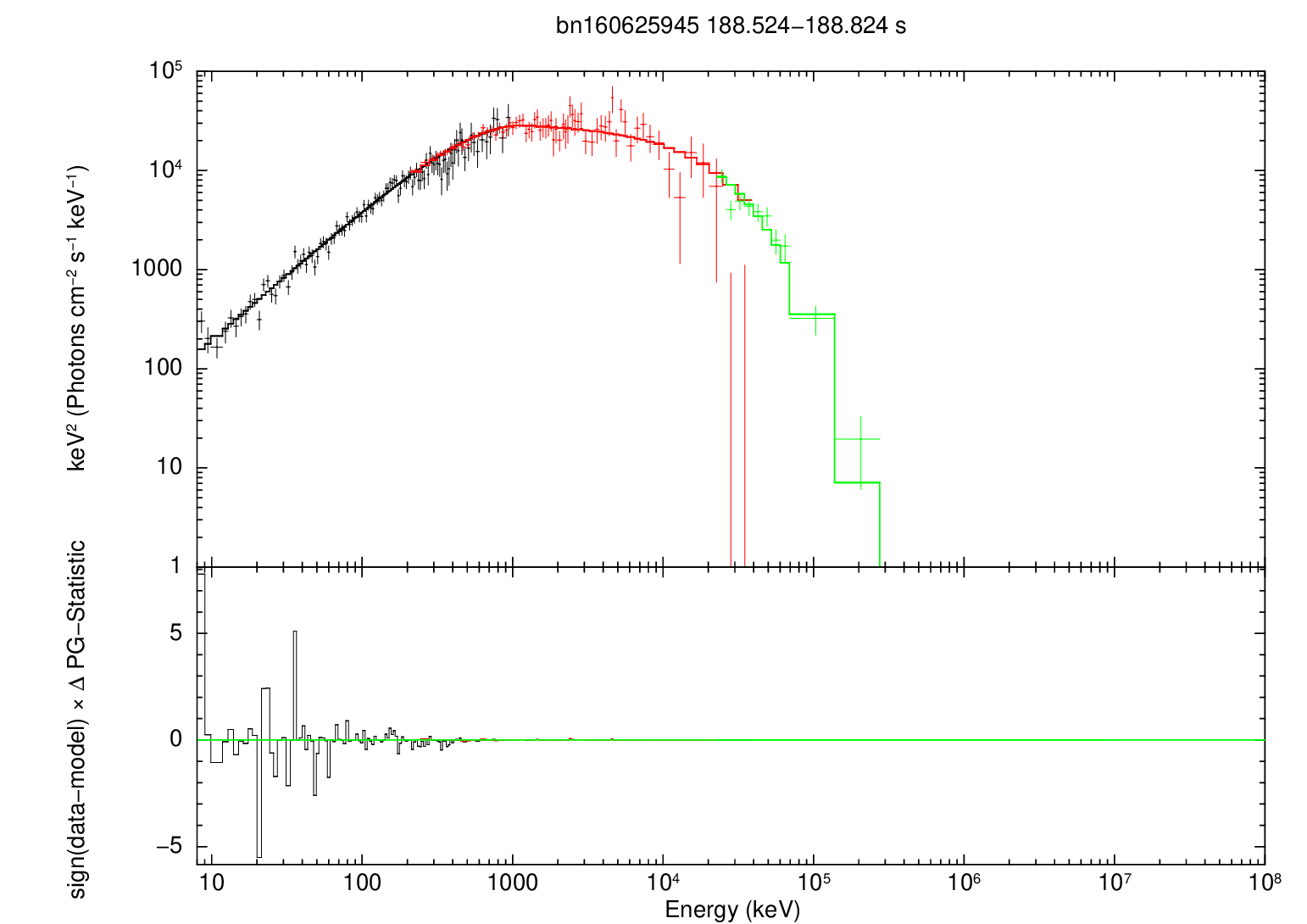} \\
\includegraphics[angle=270,scale=0.30]{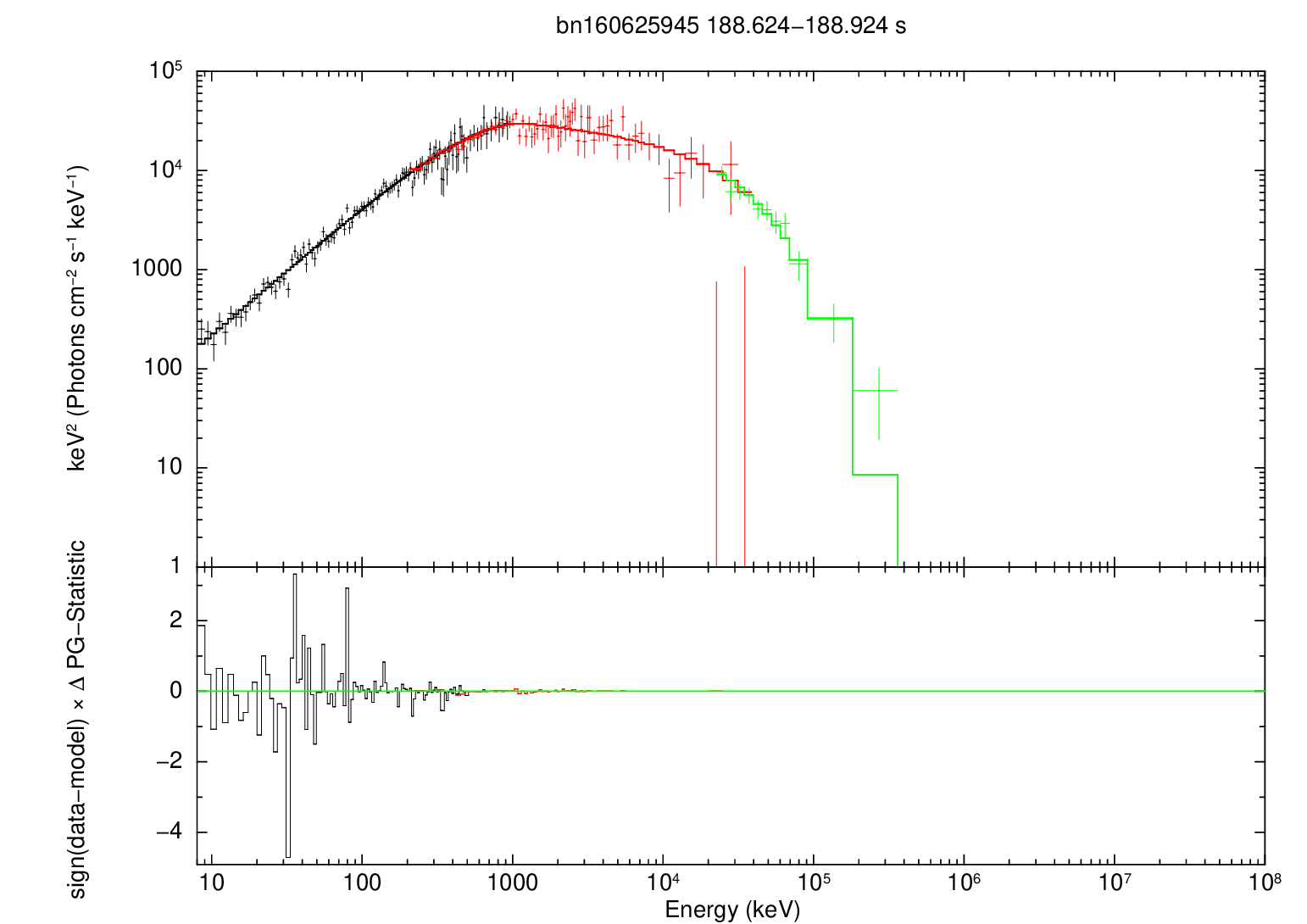} &
\includegraphics[angle=270,scale=0.30]{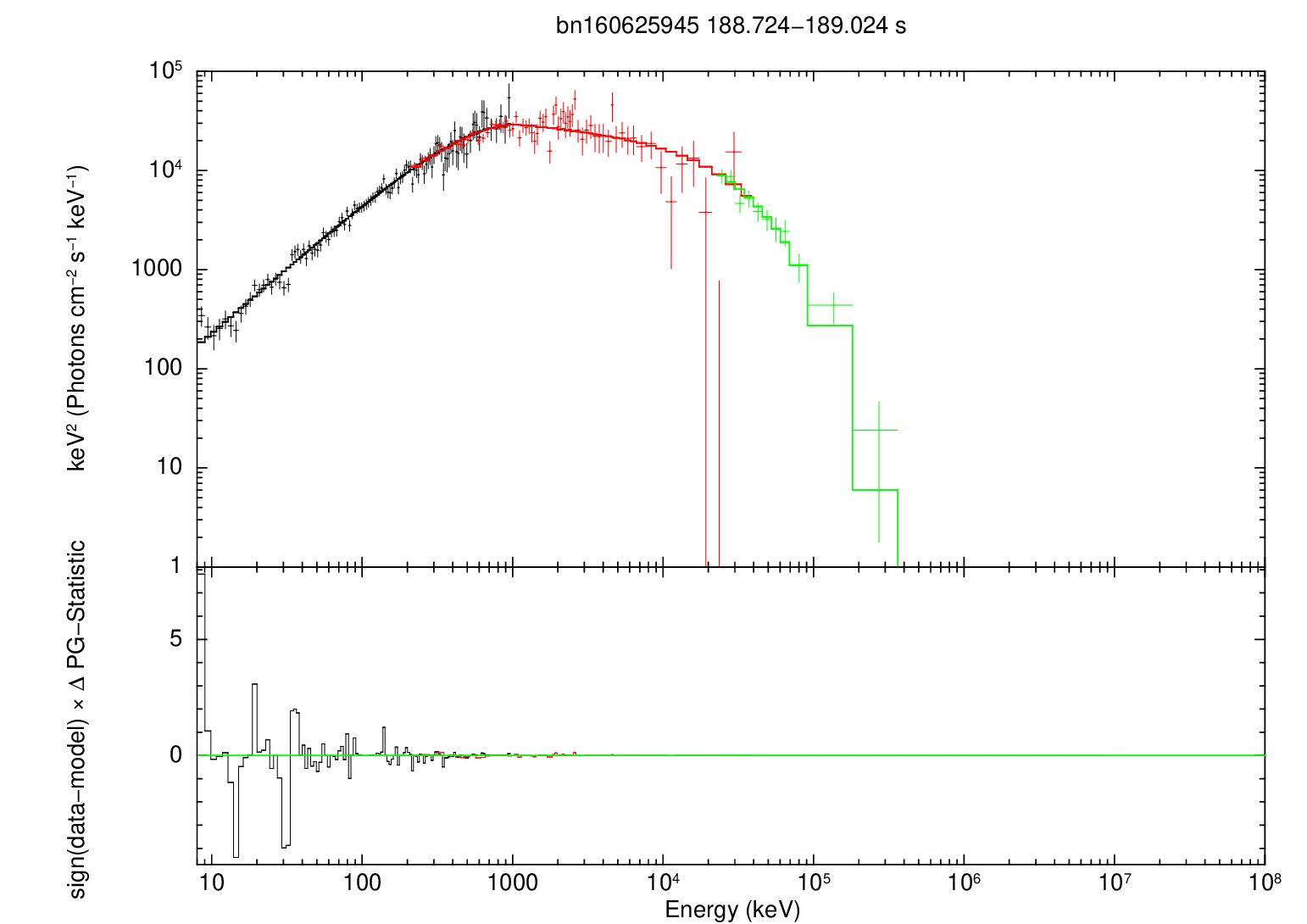} \\
\includegraphics[angle=270,scale=0.30]{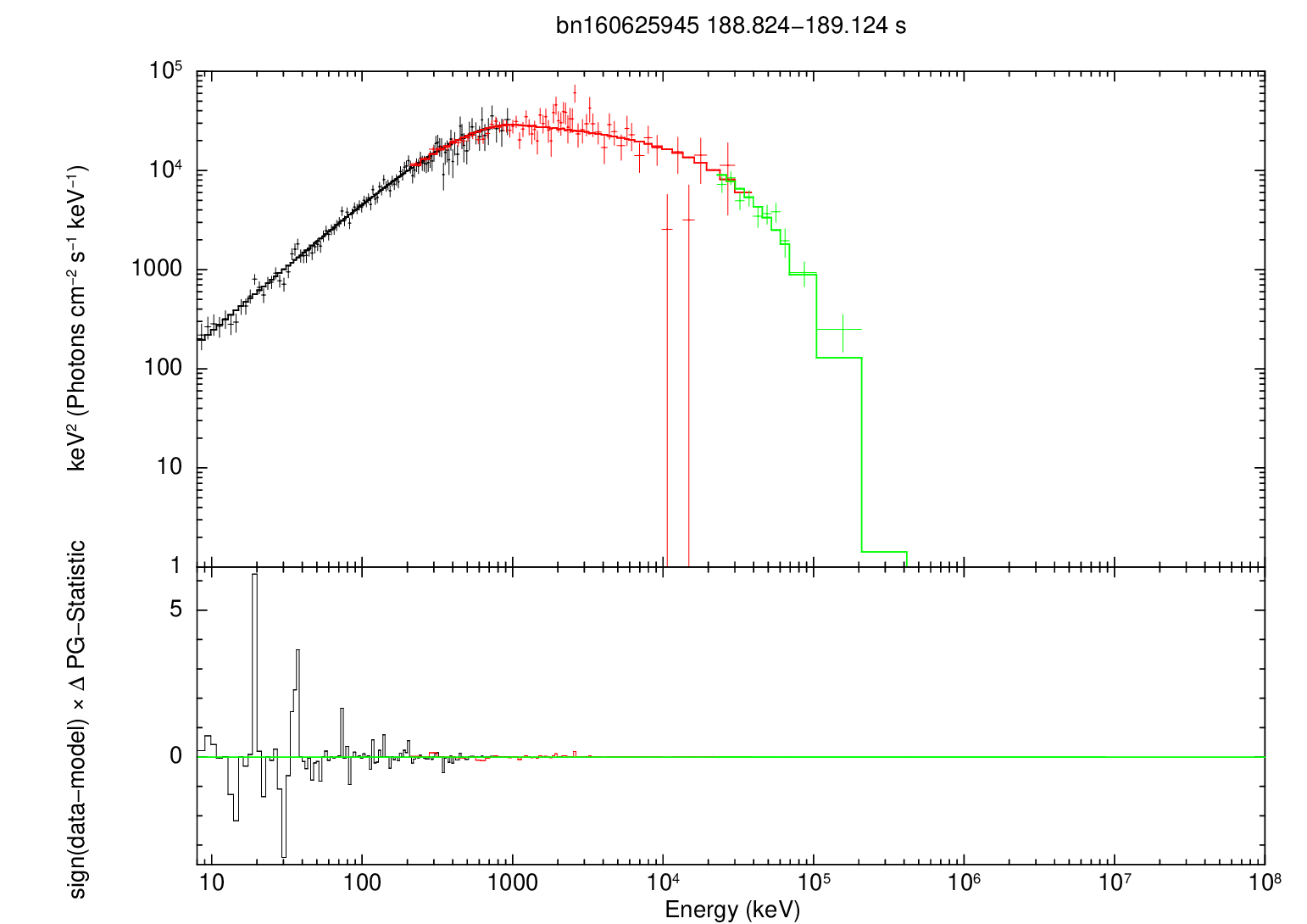} &
\includegraphics[angle=270,scale=0.30]{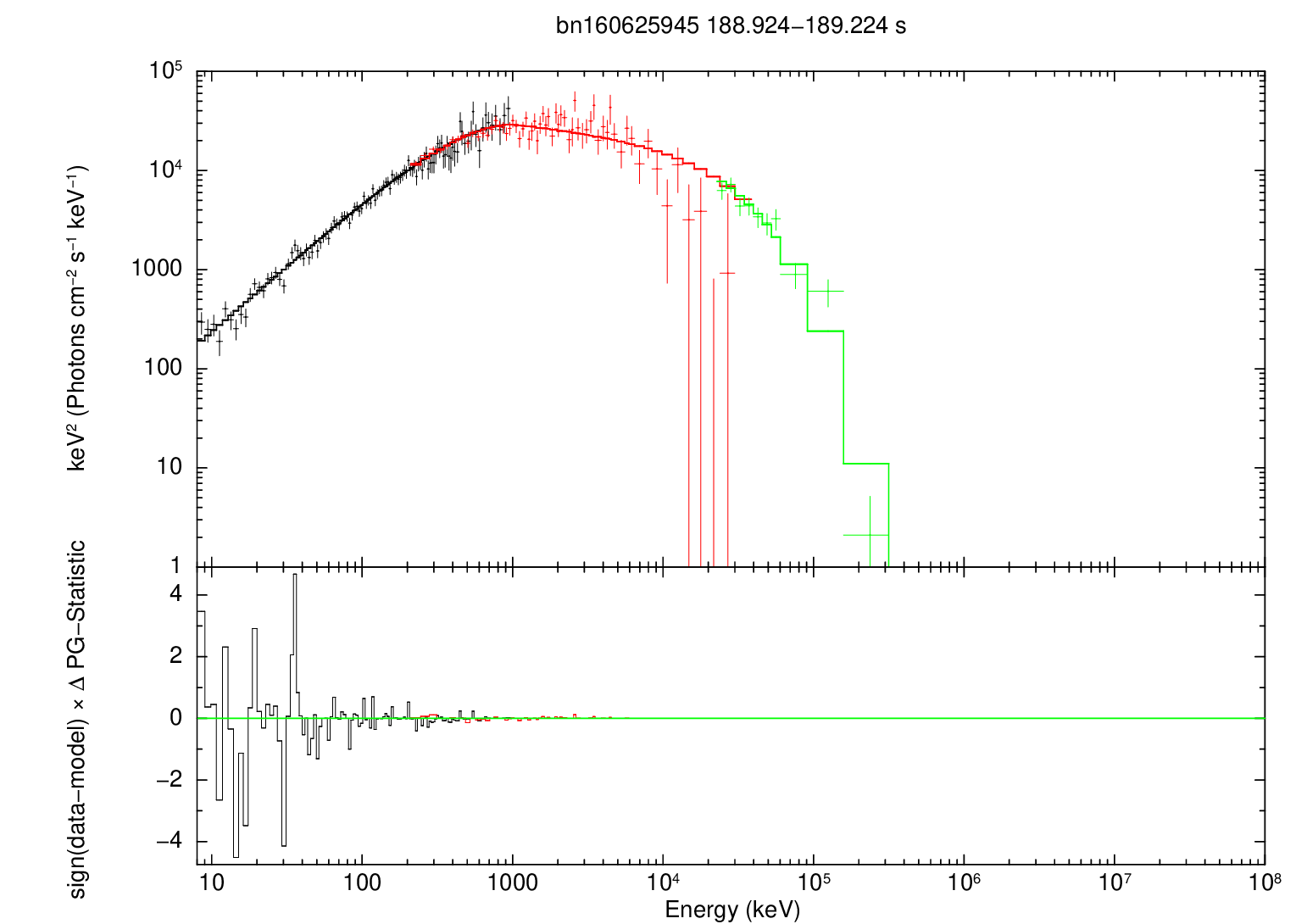} \\
\end{tabular}
\caption{(Continued)}
\end{figure*}
\addtocounter{figure}{-1}
\clearpage
\begin{figure*}
\begin{tabular}{cc}
\includegraphics[angle=270,scale=0.30]{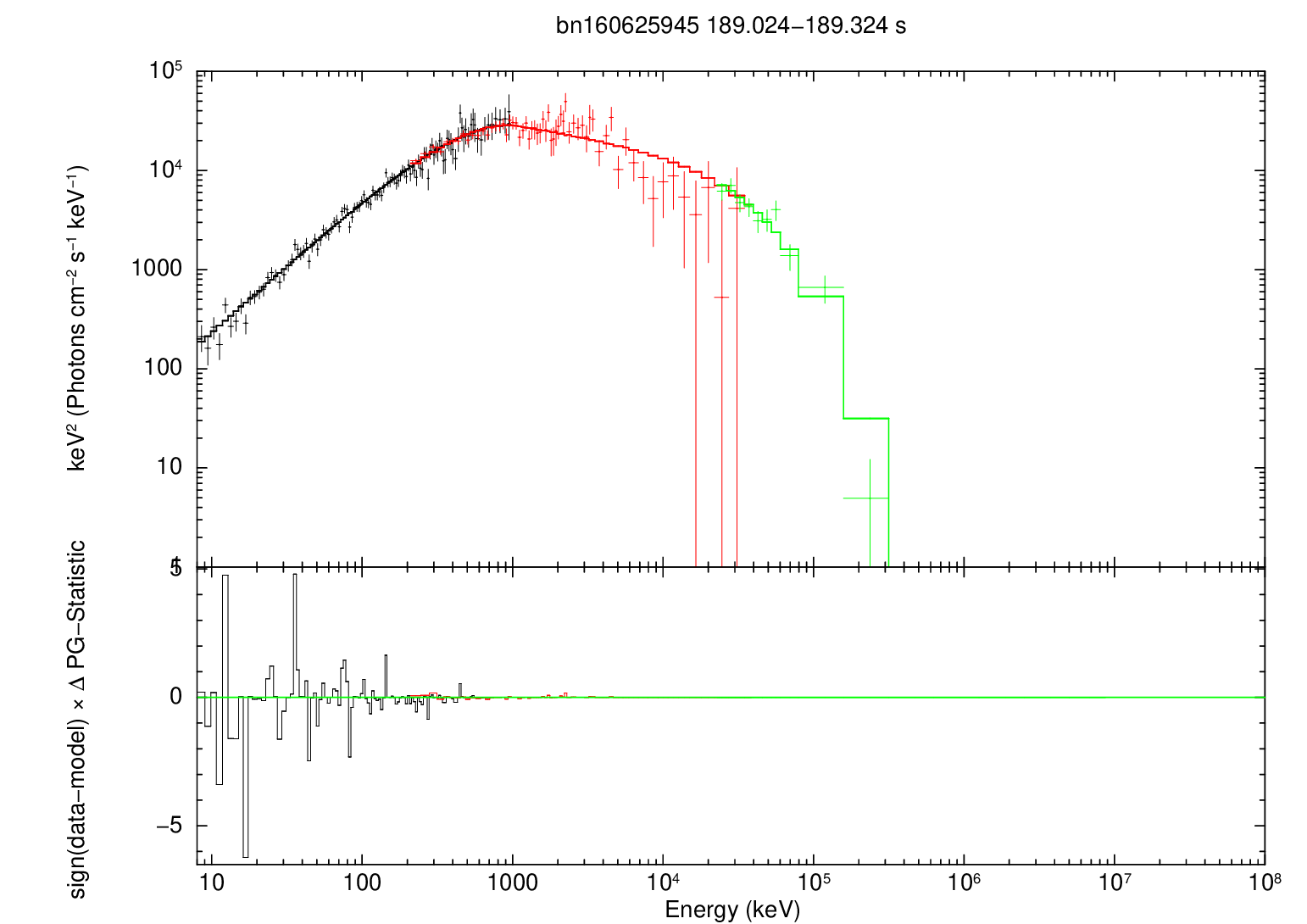} &
\includegraphics[angle=270,scale=0.30]{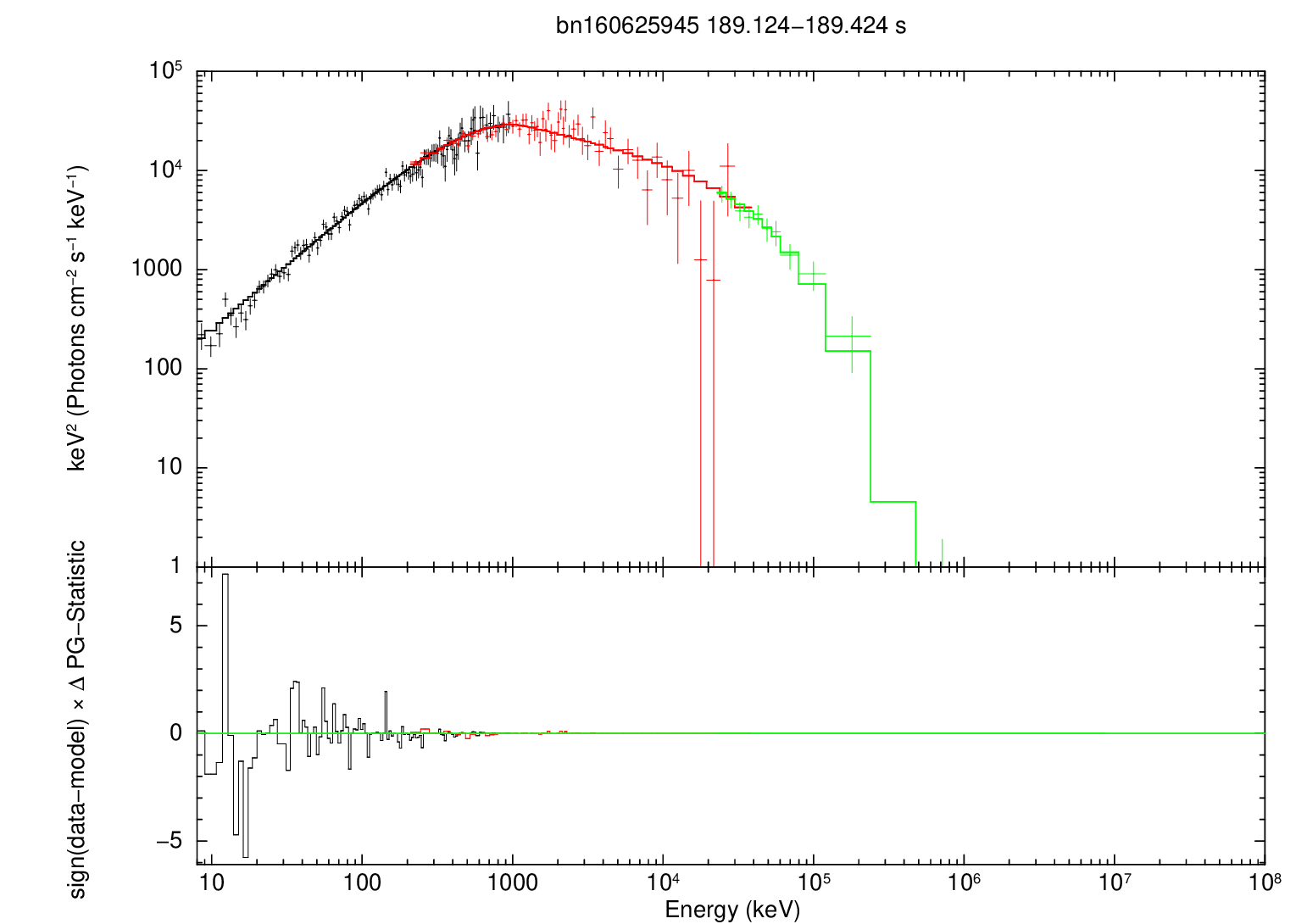} \\
\includegraphics[angle=270,scale=0.30]{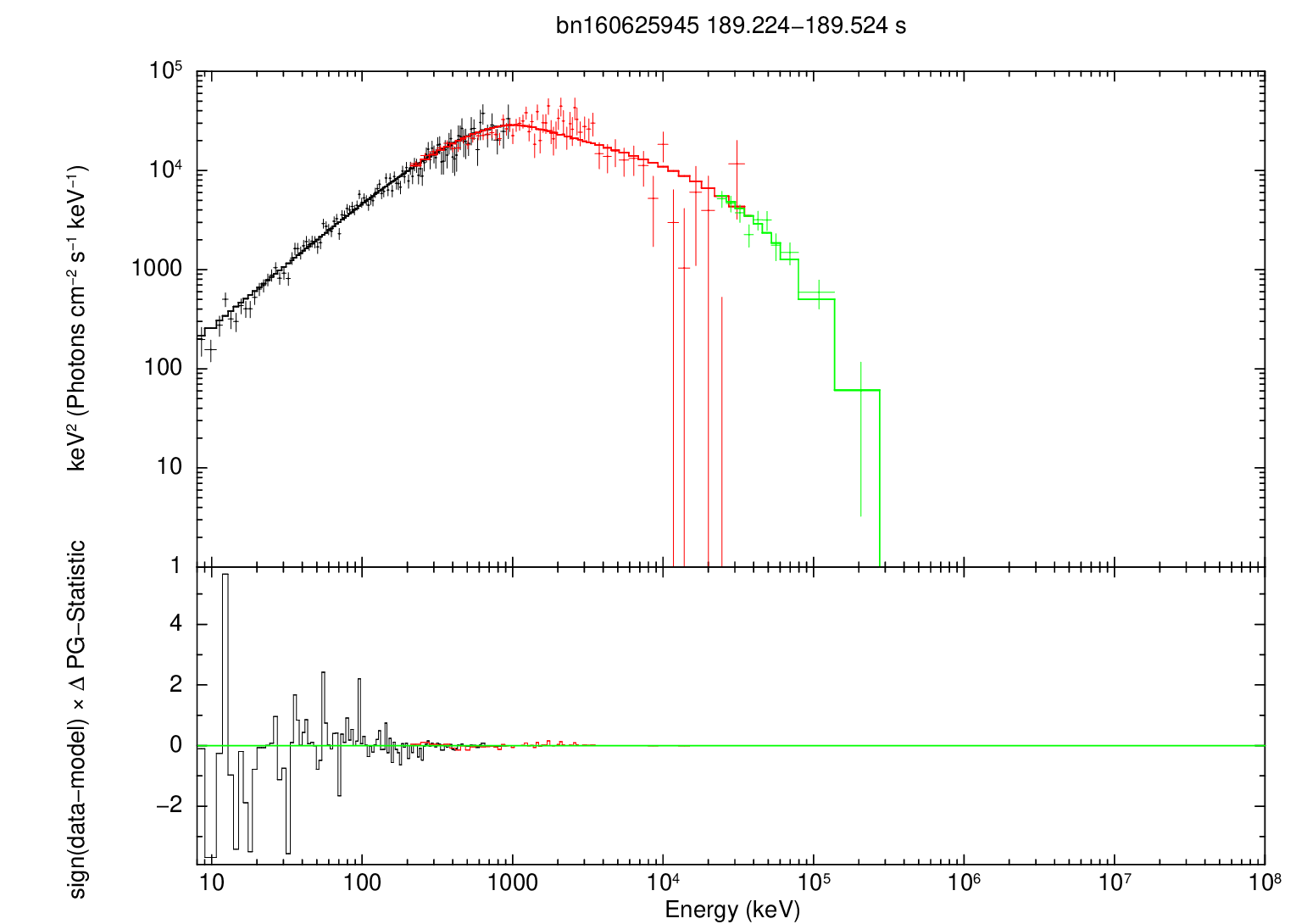} &
\includegraphics[angle=270,scale=0.30]{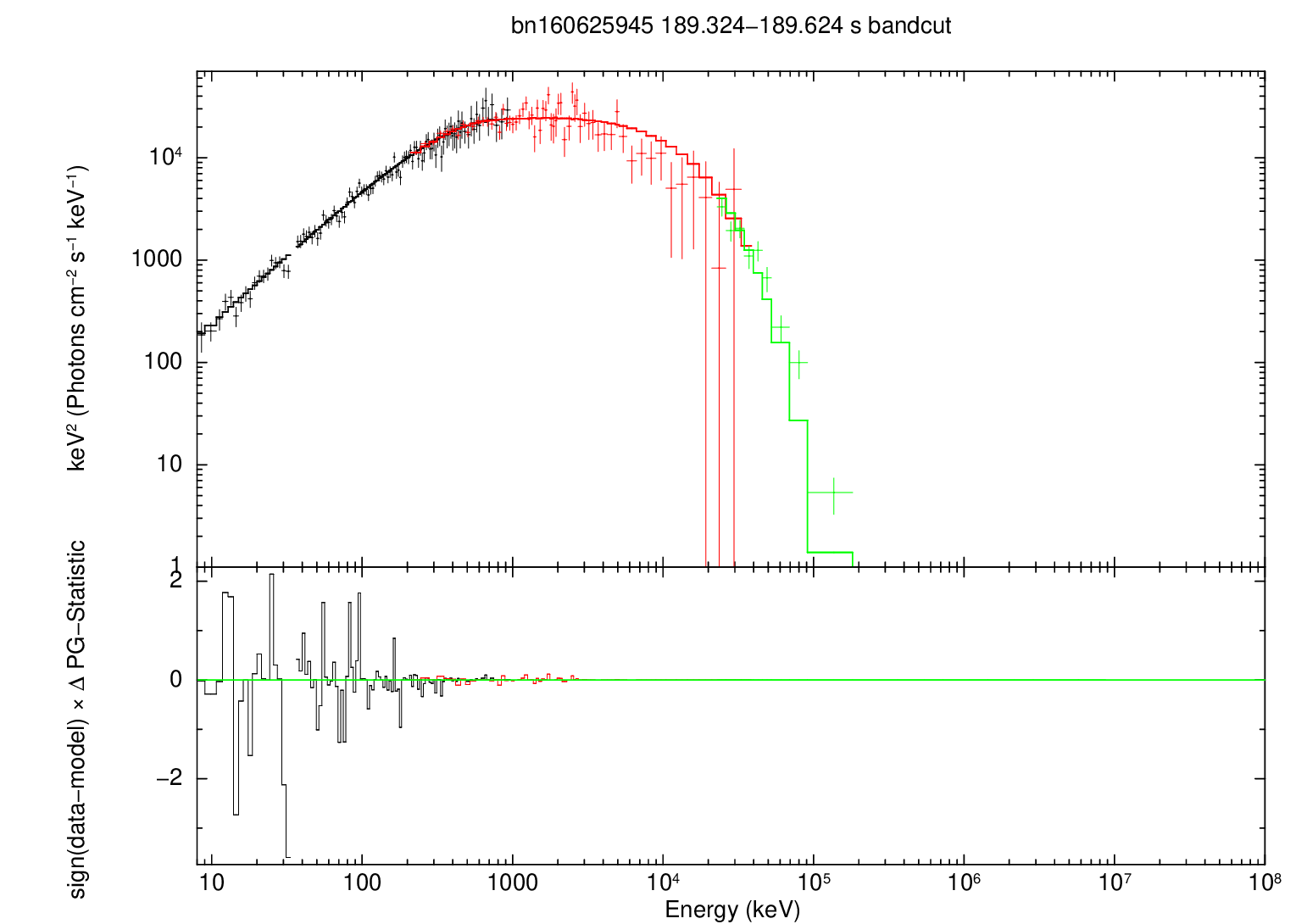} \\
\includegraphics[angle=270,scale=0.30]{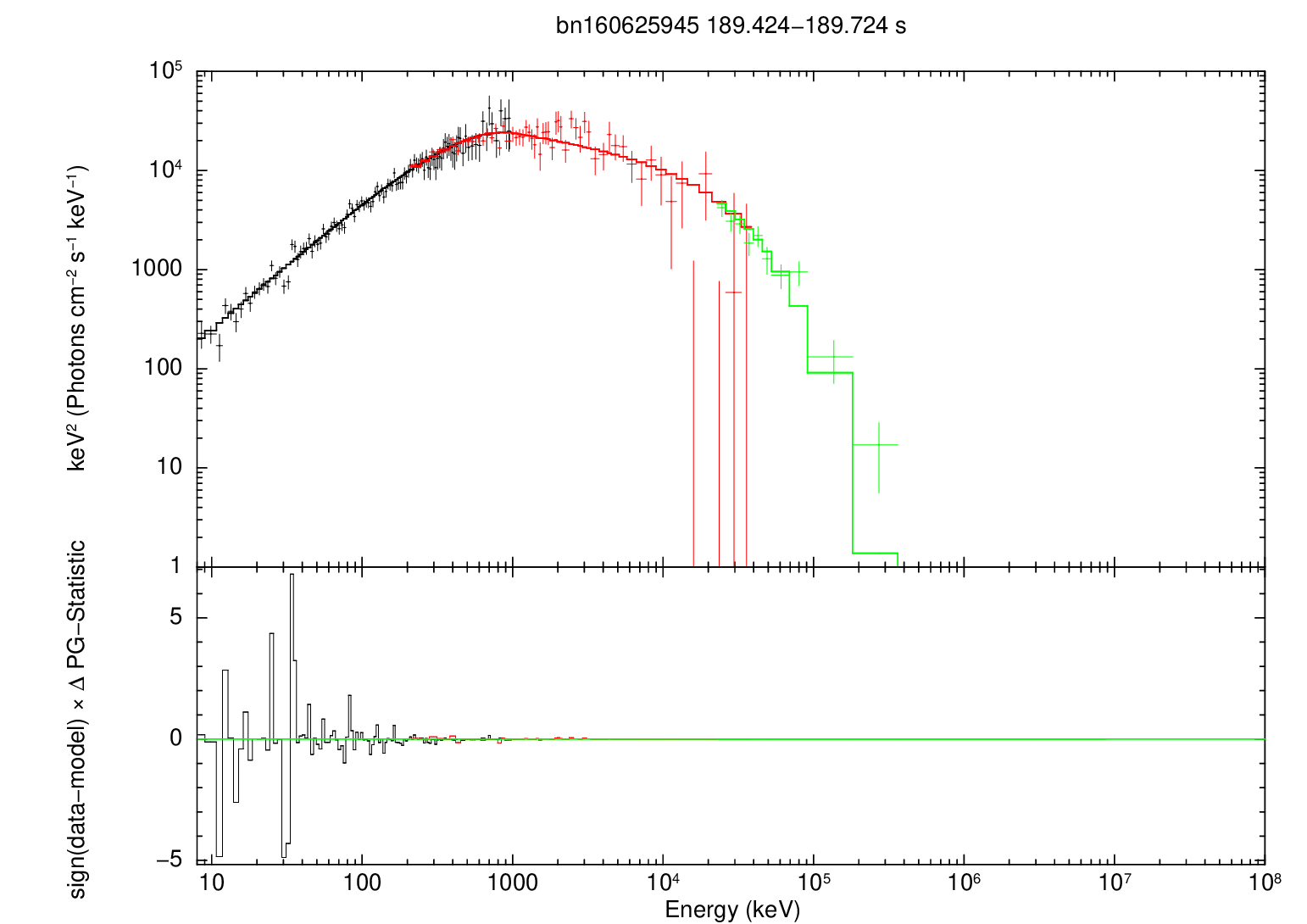} &
\includegraphics[angle=270,scale=0.30]{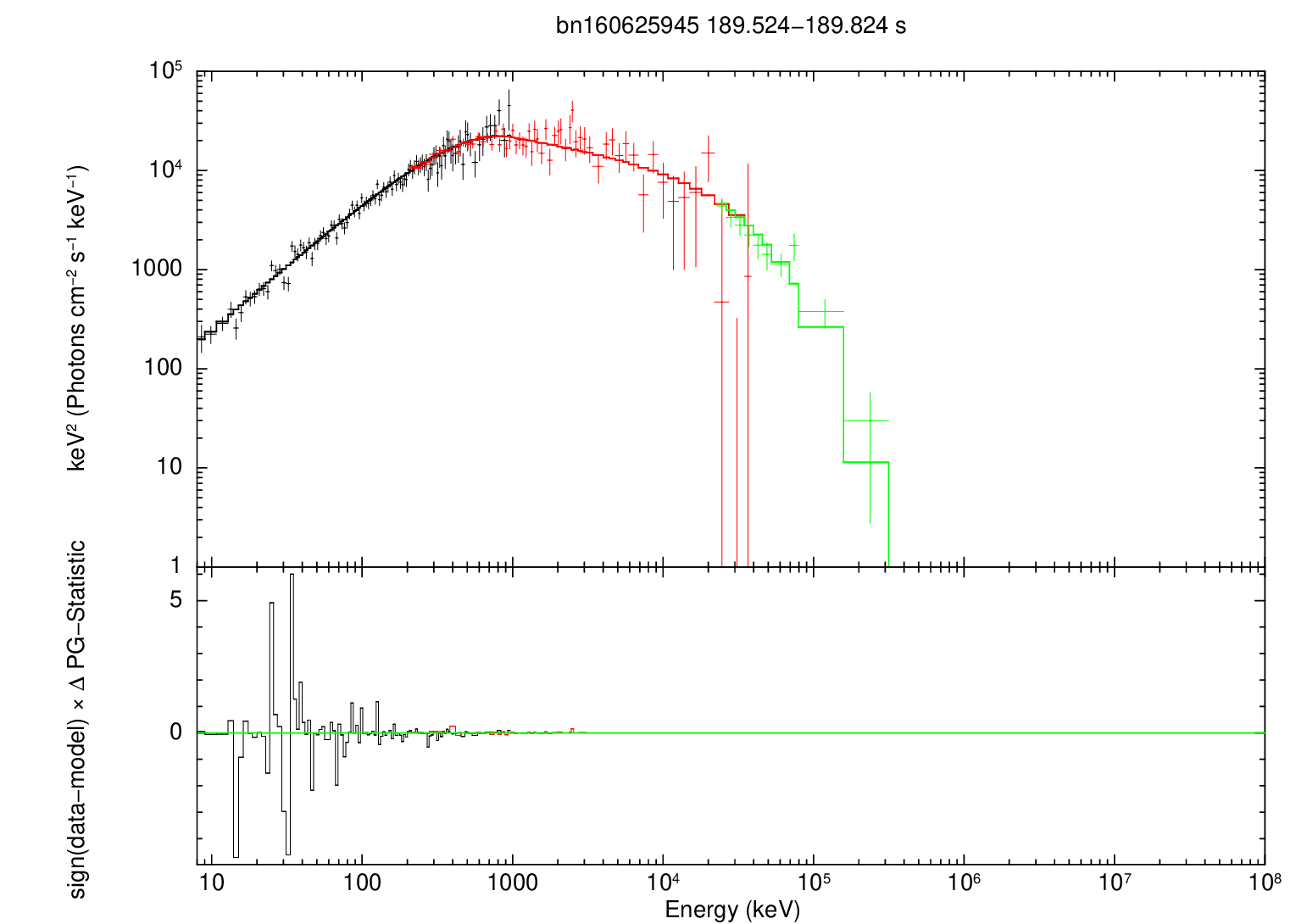} \\
\end{tabular}
\caption{(Continued)}
\end{figure*}
\addtocounter{figure}{-1}
\clearpage
\begin{figure*}
\begin{tabular}{cc}
\includegraphics[angle=270,scale=0.30]{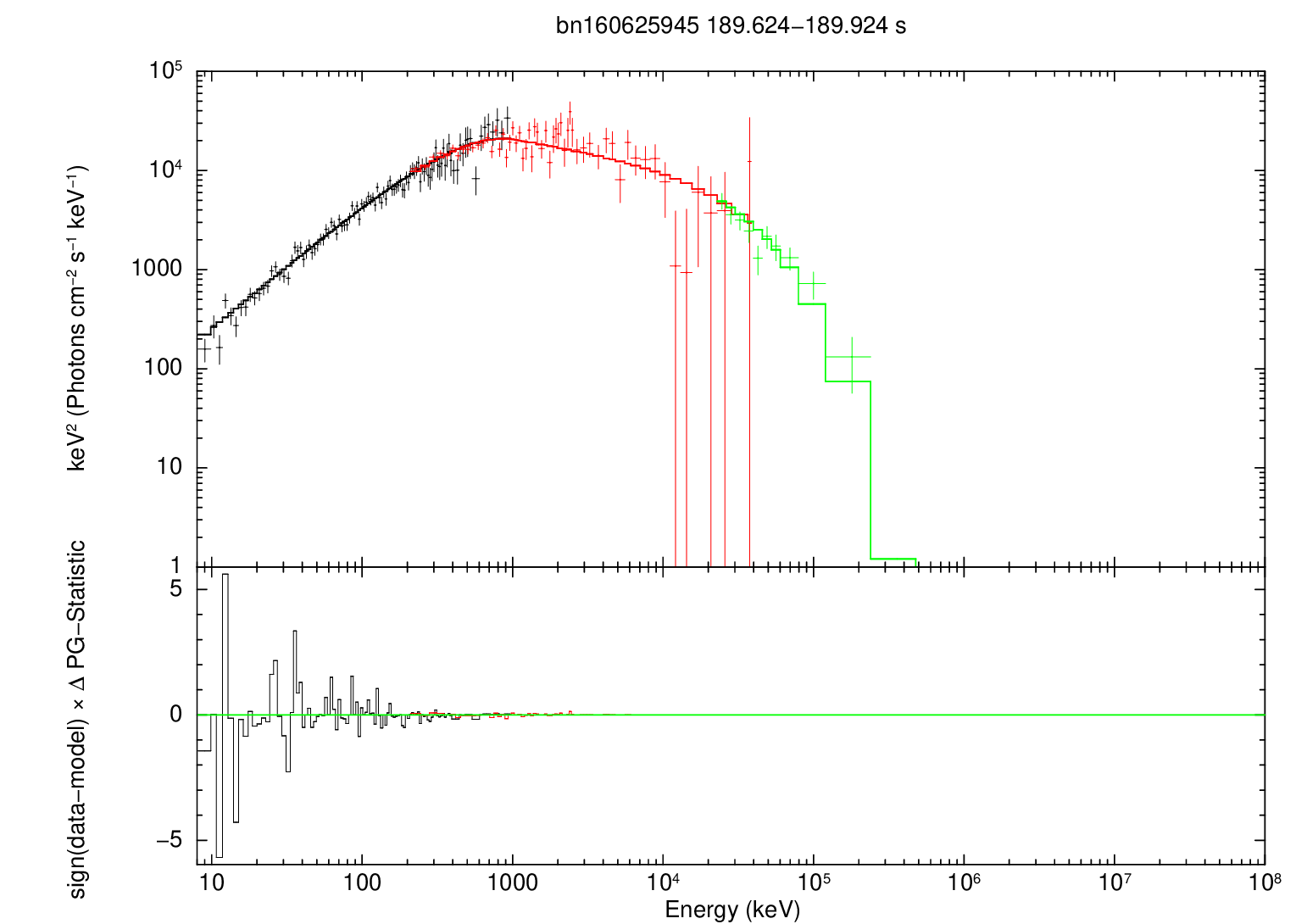} &
\includegraphics[angle=270,scale=0.30]{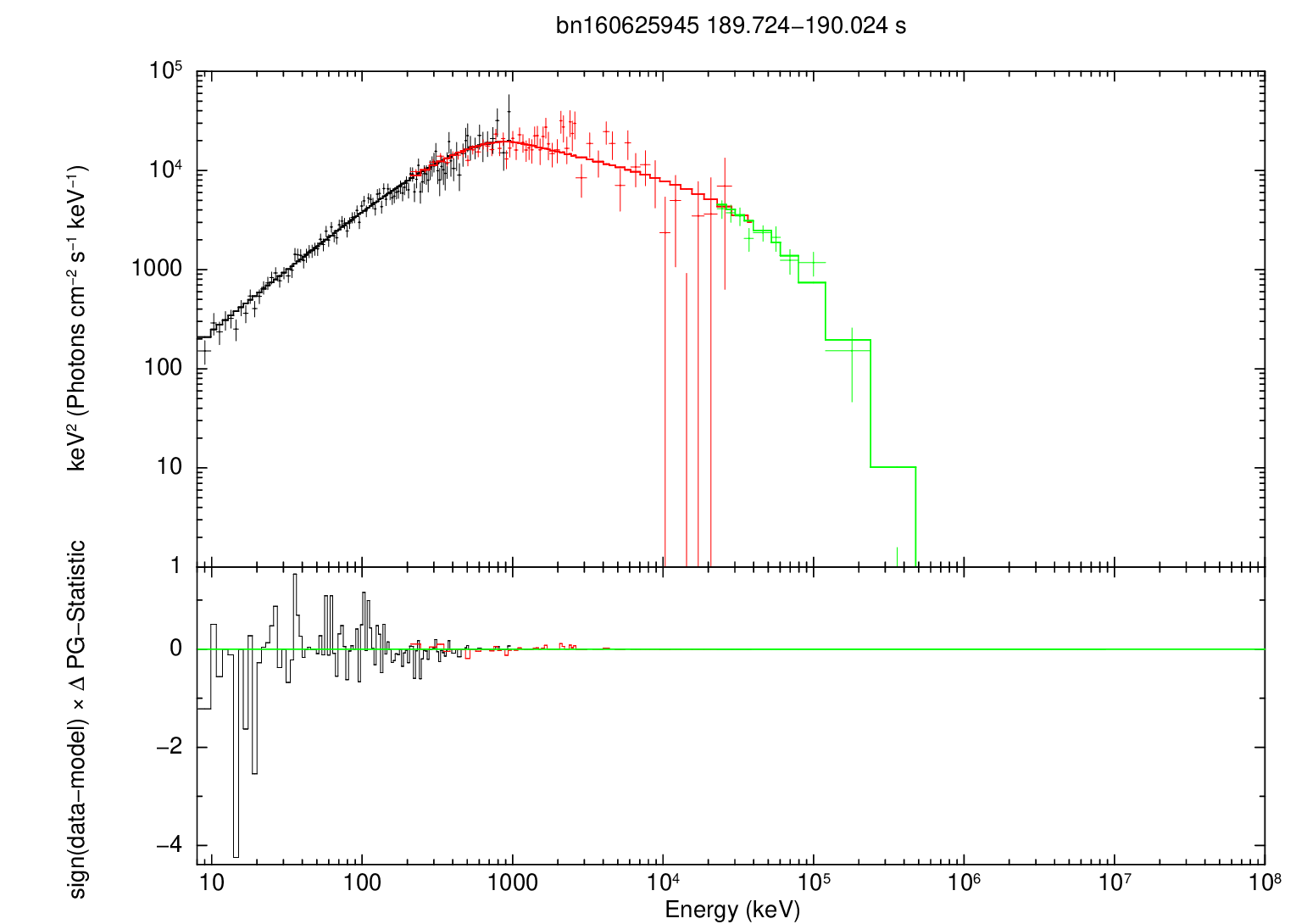} \\
\includegraphics[angle=270,scale=0.30]{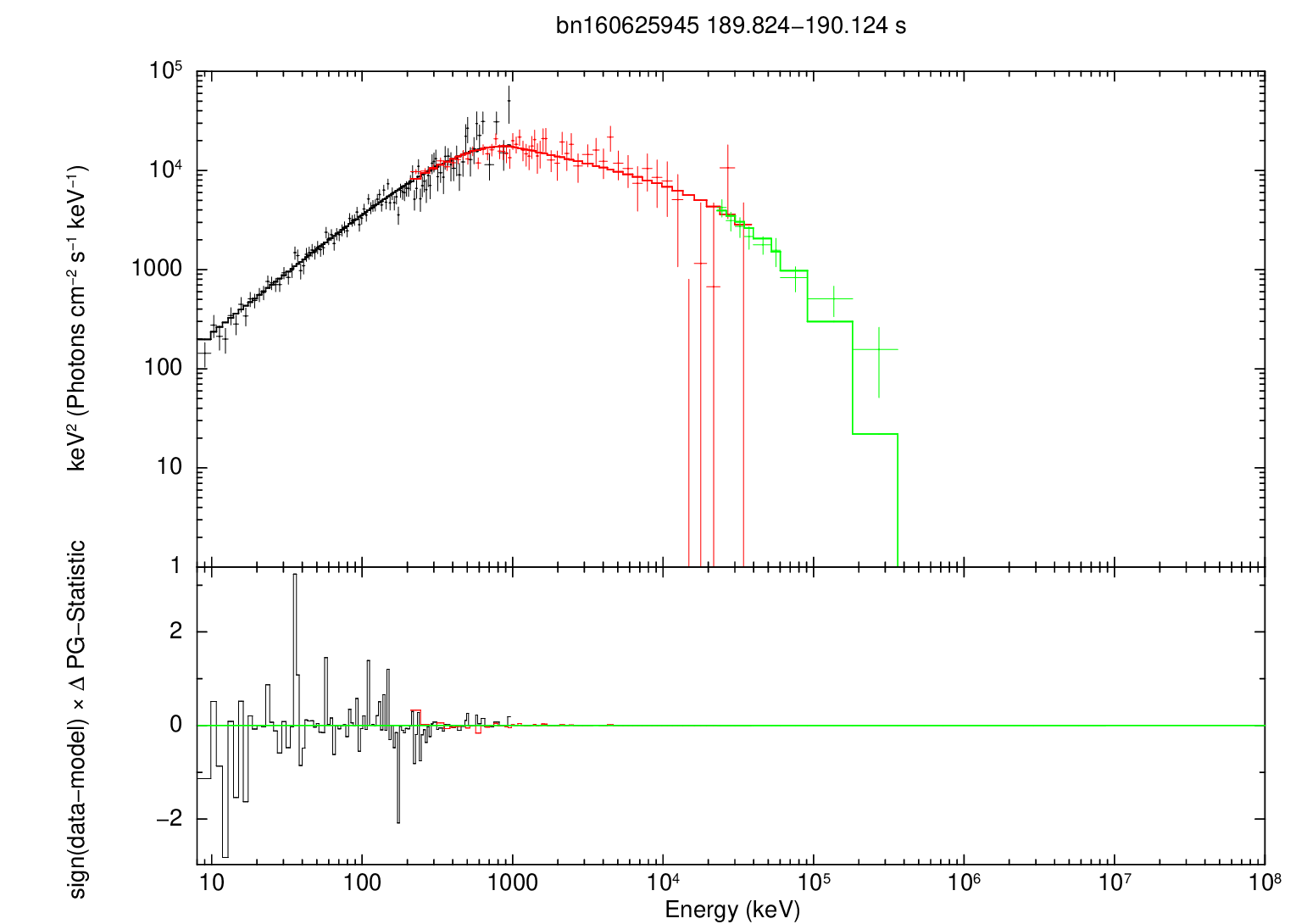} &
\includegraphics[angle=270,scale=0.30]{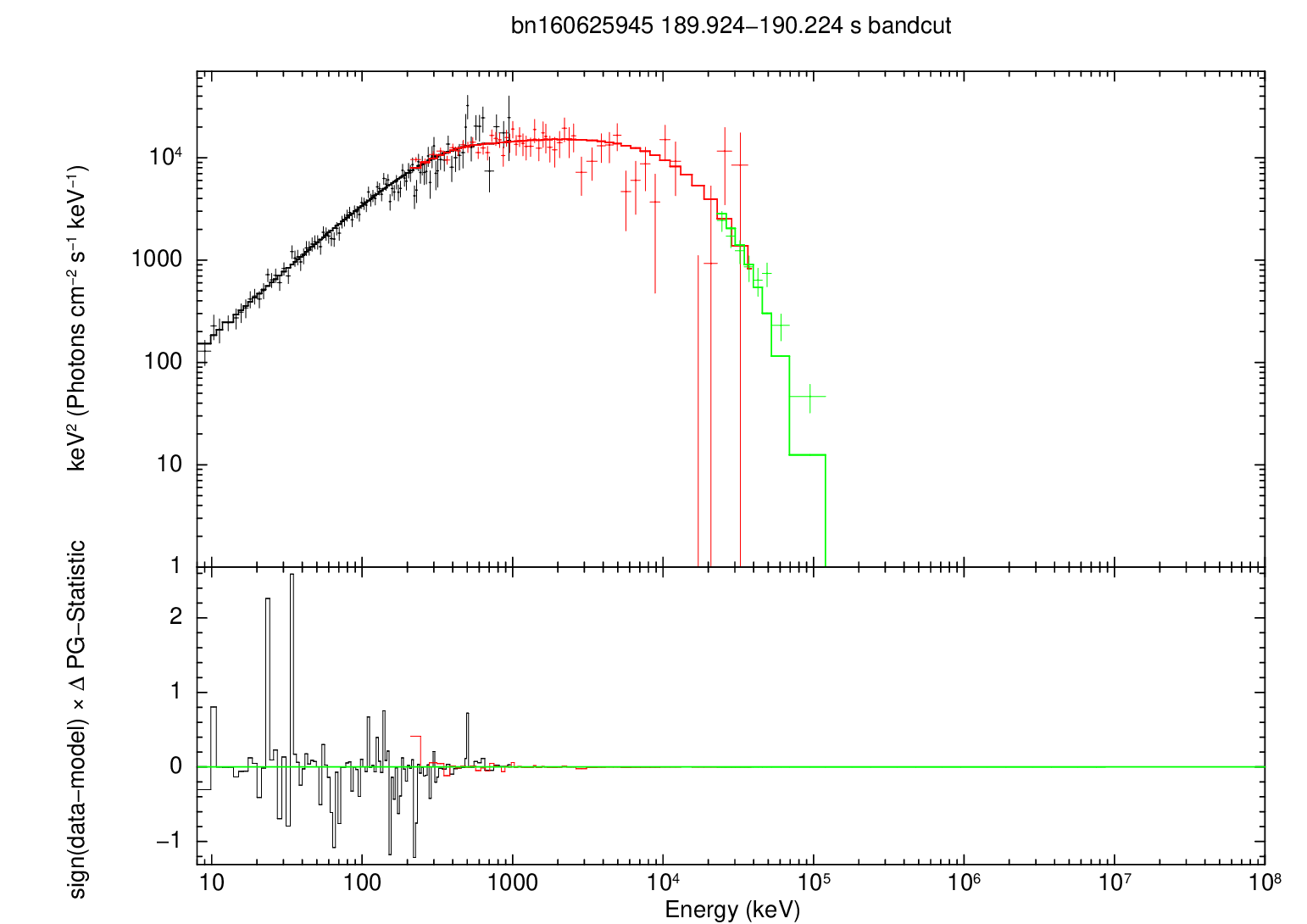} \\
\includegraphics[angle=270,scale=0.30]{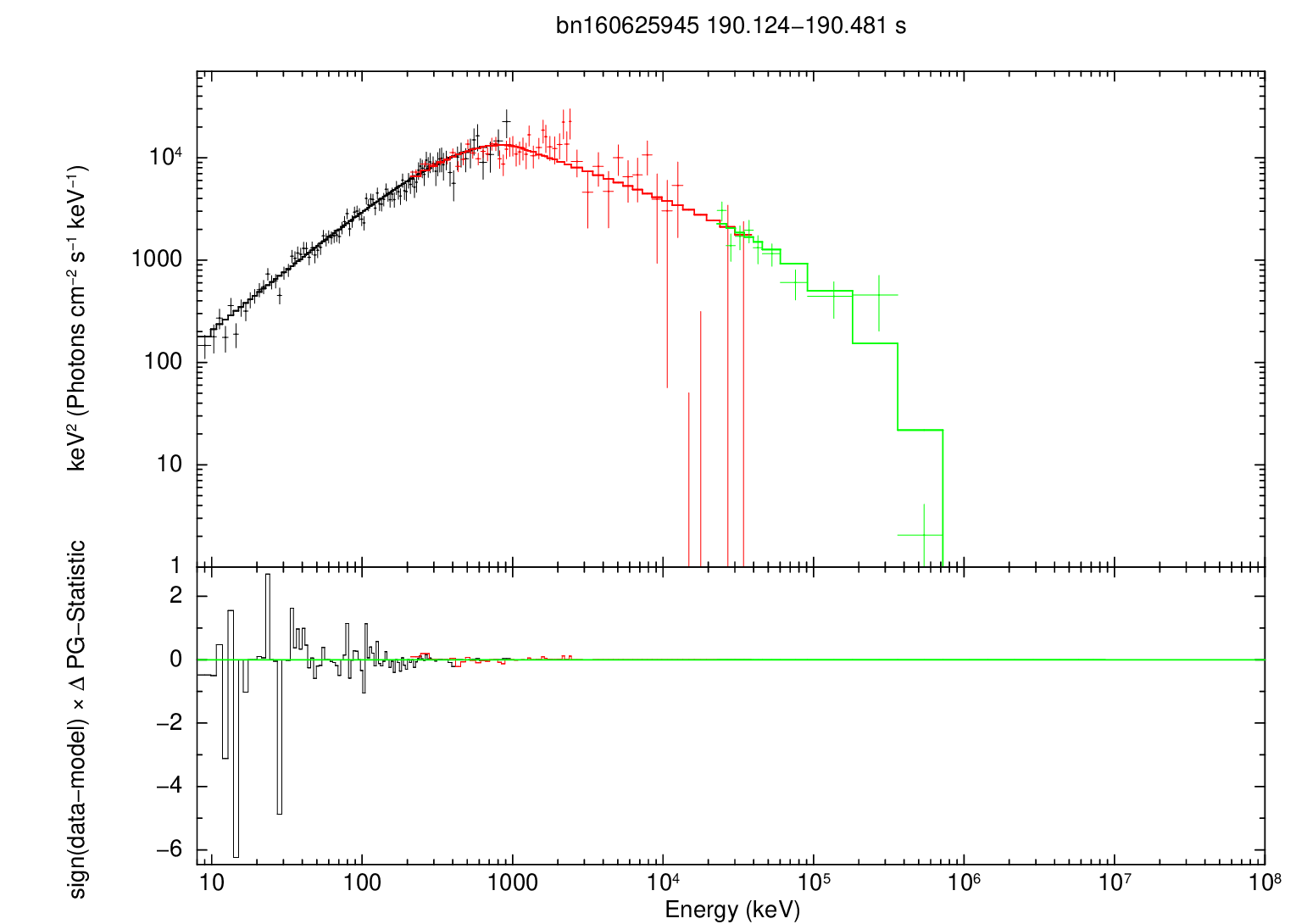}  &
\includegraphics[angle=270,scale=0.30]{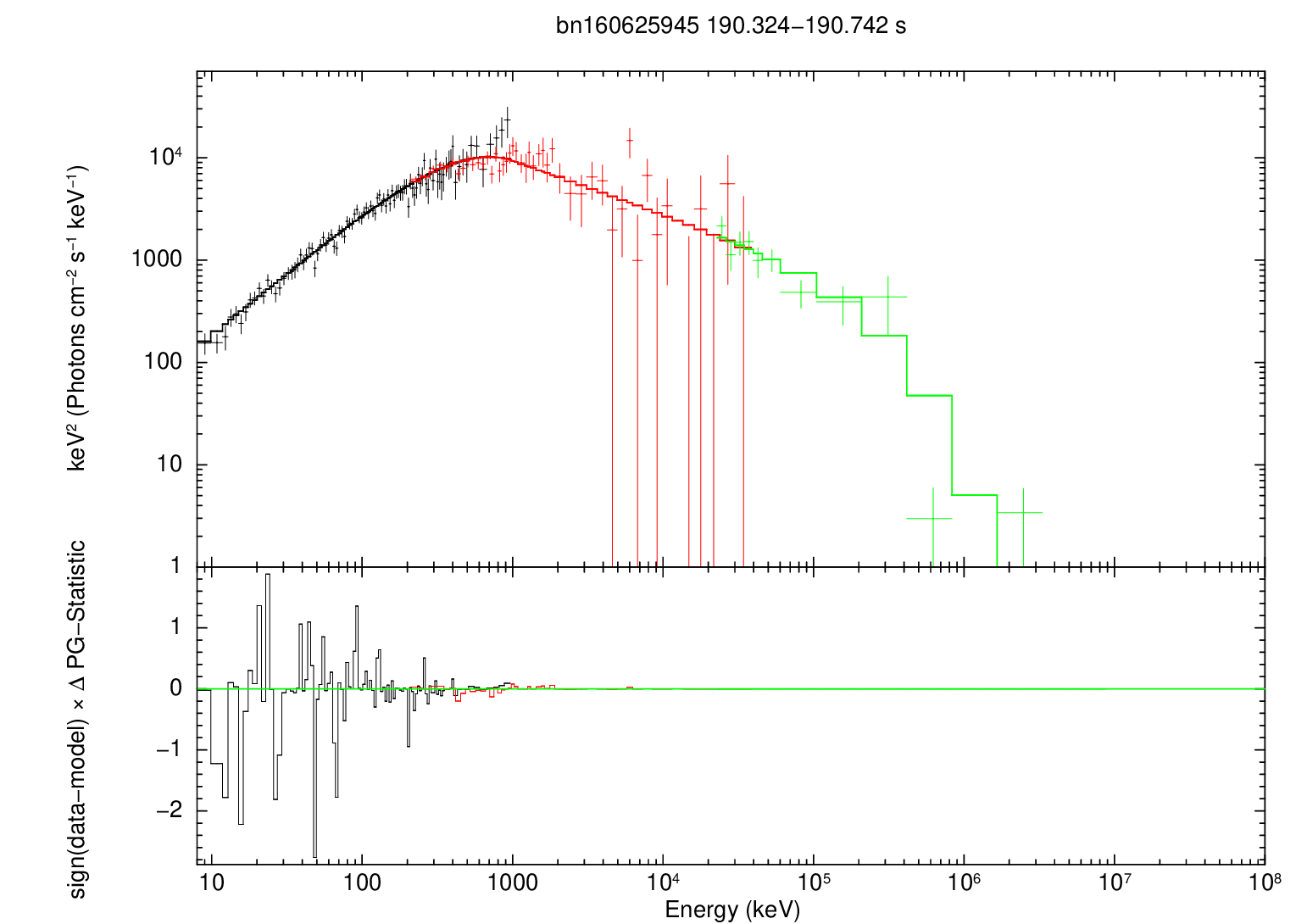}  \\
\end{tabular}
\caption{(Continued)}
\end{figure*}
\addtocounter{figure}{-1}
\clearpage
\begin{figure*}
\begin{tabular}{cc}
\includegraphics[angle=270,scale=0.30]{Fig_1_25.eps} & \\
\end{tabular}
\caption{(Continued)}
\end{figure*}

\clearpage
\begin{figure}
\centering
\includegraphics[scale=0.40]{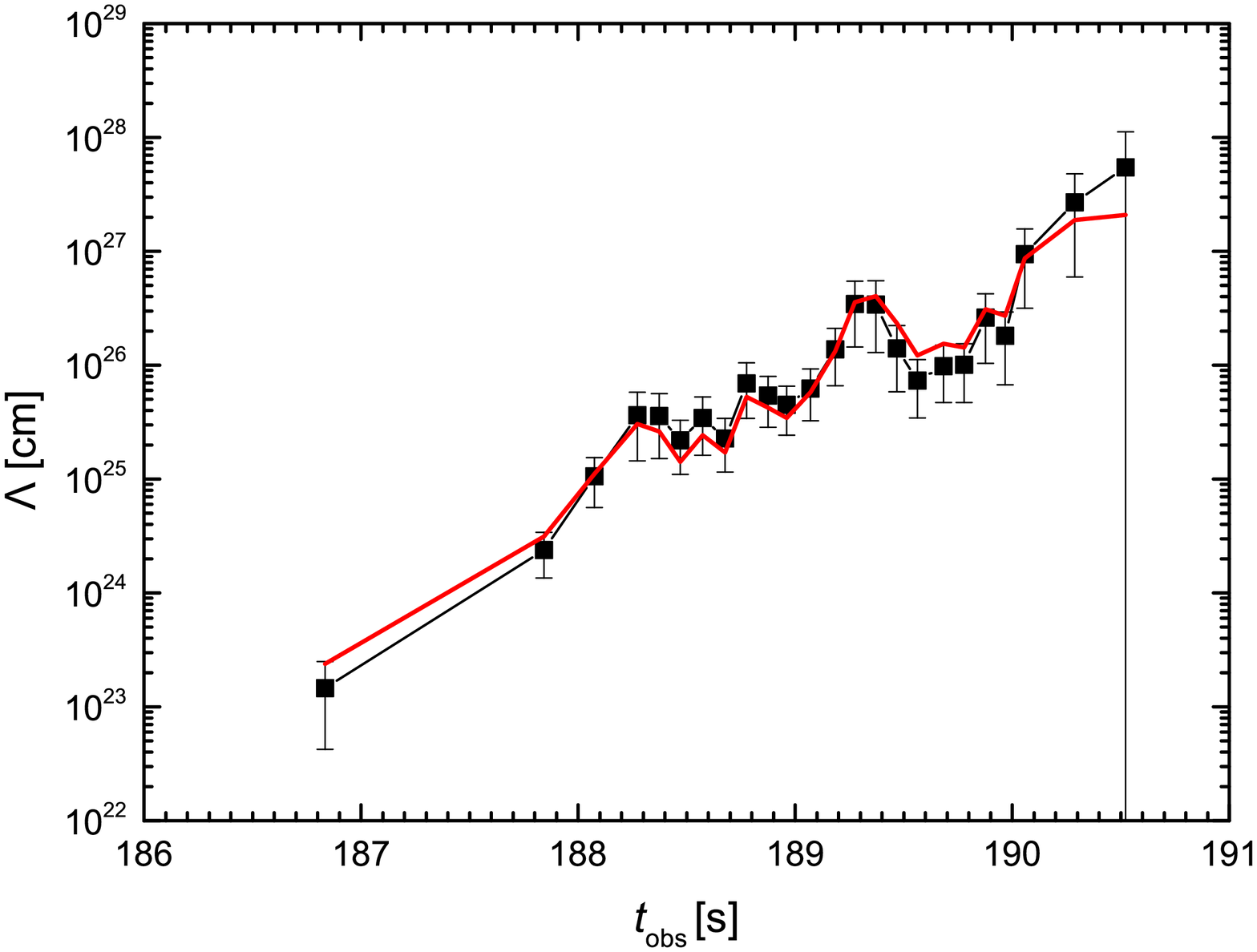}
\caption{Dependence of $\Lambda$ on $t_{\rm obs}$,
where the red line is the best fitting result by minimizing the $\chi^2$ in Equation~(\ref{Eq:chi}) with
the downhill simplex algorithm.}\label{Fig:Lambda}
\end{figure}

\clearpage
\begin{figure}
\begin{tabular}{ccc}
\includegraphics[angle=0,scale=0.28,trim=0 0 0 0,clip]{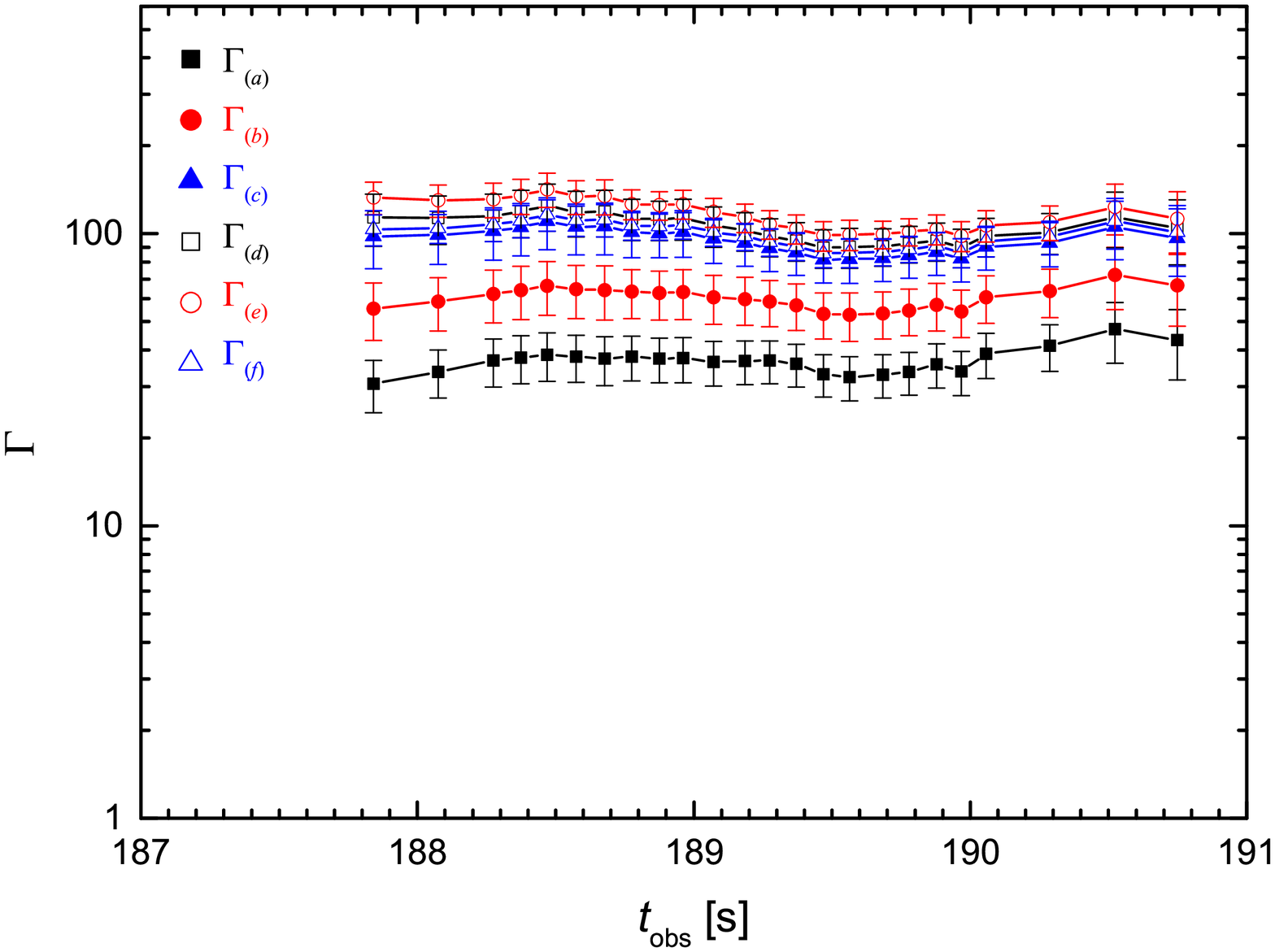} &
\includegraphics[angle=0,scale=0.28,trim=0 0 0 0,clip]{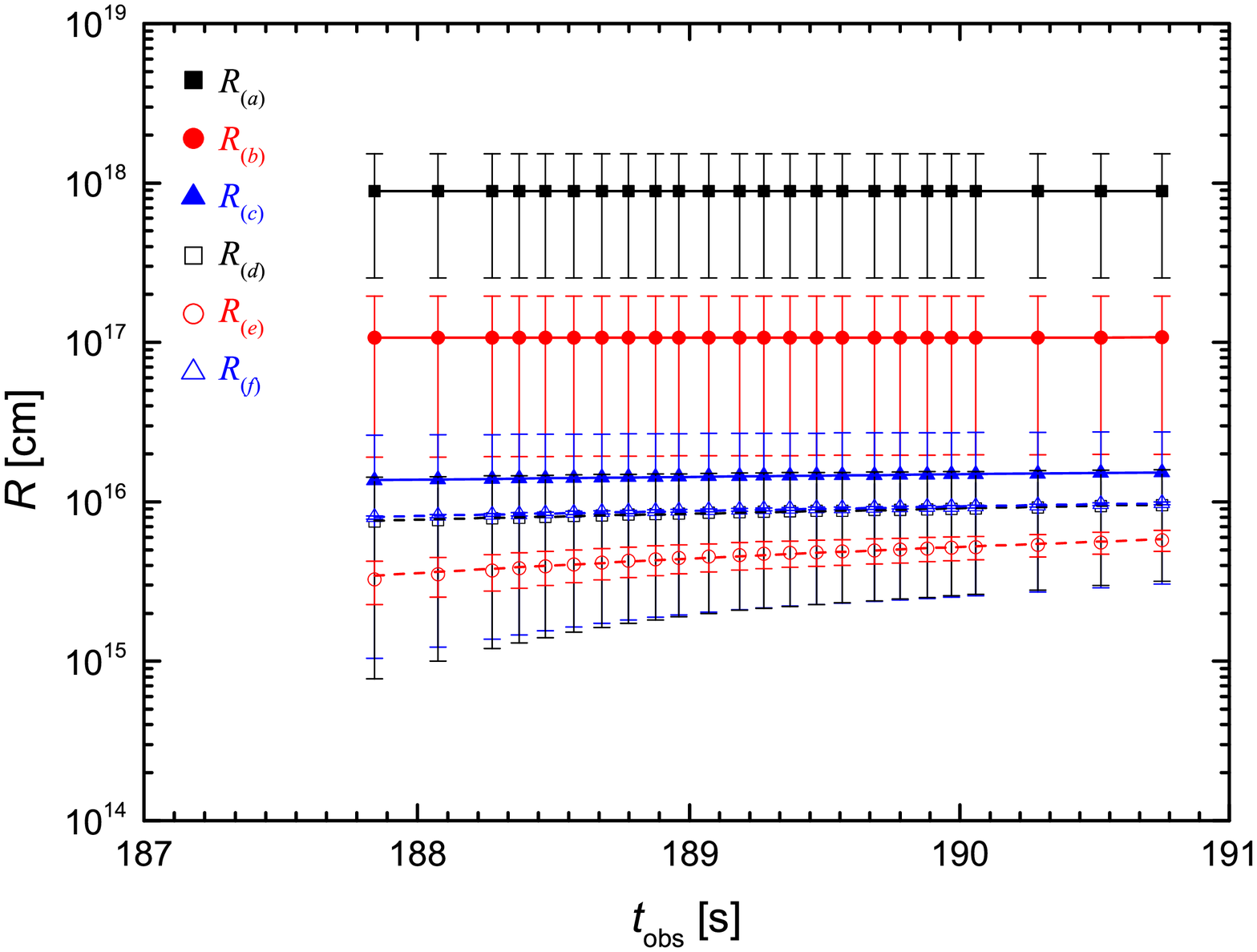}\\
\end{tabular}
\caption{Values of $\Gamma$ and $R$ at $t_{\rm obs}(\neq 186.83\rm s)$,
where the subscripts $(a)$, $(b)$, $(c)$, $(d)$, $(e)$, and $(f)$ represents
the situations with $\Gamma_{\rm try}$=25, 50, 100, 125, 250, and 500, respectively.
The lines in the right panel are the best linear fittings of $R-t_{\rm obs}$ relations,
i.e.,
$R=8.73\times 10^{17}+8.41\times 10^{13}t_{\rm obs}$ with $\chi_r^2=0.995$ for the black solid line,
$R=6.44\times 10^{16}+2.26\times 10^{14}t_{\rm obs}$ with $\chi_r^2=0.996$ for the red solid line,
$R=-8.98\times 10^{16}+5.51\times 10^{14}t_{\rm obs}$ with $\chi_r^2=0.992$ for the blue solid line,
$R=-1.19\times 10^{17}+6.74\times 10^{14}t_{\rm obs}$ with $\chi_r^2=0.989$ for the black dashed line,
$R=-1.50\times 10^{17}+8.18\times 10^{14}t_{\rm obs}$ with $\chi_r^2=0.985$ for the red dashed line,
and
$R=-1.05\times 10^{17}+6.01\times 10^{14}t_{\rm obs}$ with $\chi_r^2=0.992$ for the blue dashed line,
respectively.
}\label{Fig:R_Gamma}
\end{figure}

\clearpage
\appendix
\section{Joint Spectral Fitting with/without LAT}\label{Appendix:A}
In this work, we perform joint spectral fittings of the NaI, BGO, and LLE data,
of which the obtained results are used to estimate the $\Gamma$ and $R$ of an expanding jet shell.
We would like to point out that it is safe to use the LLE data
in the spectral analysis for the FP2EE in GRB~160625B.
The reasons are shown as follows.

\begin{figure*}
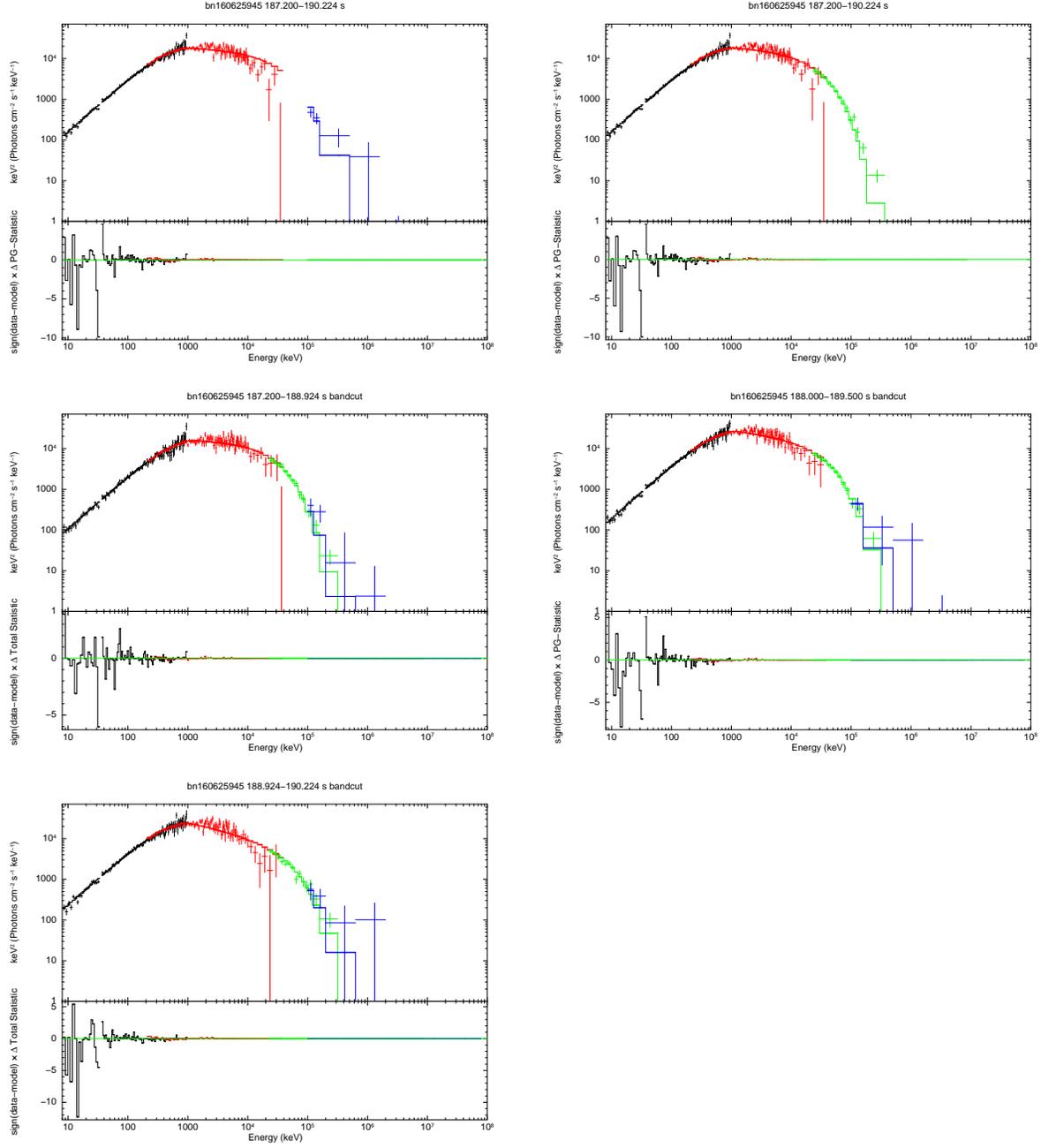

\centering
\begin{tabular}{cc}
\includegraphics[angle=270,scale=0.30]{Fig_1_81_LAT.eps} &
\includegraphics[angle=270,scale=0.30]{Fig_1_80_LLE.eps} \\
\includegraphics[angle=270,scale=0.30]{Fig_1_60_LAT.eps} &
\includegraphics[angle=270,scale=0.30]{Fig_1_61_LAT.eps} \\
\includegraphics[angle=270,scale=0.30]{Fig_1_62_LAT.eps} & \\
\end{tabular}
\caption{Same as Figure~2,
where the LAT data is shown with blue ``$+$''.}\label{Fig:Joint_Spectral_fittings_LAT}
\end{figure*}

We perform a joint spectral fitting of the NaI, BGO, and LAT/LLE data for the FP2EE of GRB~160625B.
Here, we use the LAT Pass~8 data, which is reduced by using the ScienceTools-v10r0p5-fssc-20150518A-source package
and the P8R2\_TRANSIENT020E\_V6 response function\footnote{For detailed information about the LAT GRB analysis,
please see the NASA Fermi Web site.}.
The fitting is shown in the left/right-top panel of Figure~\ref{Fig:Joint_Spectral_fittings_LAT}
and the obtained result is reported in the second/third row of Table~\ref{Appendix:TabLAT}.
By comparing the values in the second row with those in the third row of Table~\ref{Appendix:TabLAT},
one can conclude that the result from the NaI+BGO+LLE joint spectral fitting
is consistent with that from the NaI+BGO+LAT joint spectral fitting.
Since the number of the observed LAT photons is low,
the NaI+BGO+LAT joint spectral fitting is only carried out in the time interval $[187.20,190.22]$~s rather than a shorter time interval.
We also perform joint spectral fittings of the NaI, BGO, LLE, and LAT data in three time intervals,
i.e., $[187.20,188.92]$~s, $[188.00,189.50]$~s, and $[188.92,190.22]$~s.
The fittings are shown in Figure~\ref{Fig:Joint_Spectral_fittings_LAT} and
the obtained results are reported in Table~\ref{Appendix:TabLAT}.
One can easily find that the LLE data smoothly connect with the BGO data at $\sim$20~MeV
(see also Figure~\ref{Fig:Joint_Spectral_fittings_LLE})
and the LAT data at $\sim$100~MeV.
By comparing the results in Table~\ref{Appendix:TabLAT} with those in Table~\ref{Tab:Fitting_Results},
the results from the NaI+BGO+LLE spectral fitting are consistent with those from the NaI+BGO+LLE+LAT spectral fitting.
Then, it is safe to use the LLE data in the spectral analysis for the FP2EE in GRB~160625B.
It should be noted that the high value of $\chi_r^2$ reported in Table~\ref{Appendix:TabLAT}
is owing to the strong spectral evolution, which can be found in Table~\ref{Tab:Fitting_Results}.

\clearpage

\clearpage
\begin{table}\label{Appendix:TabLAT}
\caption{Spectral fitting results of the FP2EE in GRB~160625B with/without LAT data.}
\label{Tab:LAT_Fittings}
\begin{tabular}{c|cccccccc}
\hline\hline
Data  &	Time Interval (s) &	$\alpha$ &	$\beta$			&	$E_0  ({\rm keV})$			&	$E_1 ({\rm MeV})$			&	$N_0$\tablenotemark{a}			&	$\Lambda ({\rm cm})$			&	$\chi_r^{2}$	\\
\hline																					
NaI+BGO+LAT	&$[	187.20 	,	190.22 	]$&$	-0.68 	\pm	0.01 	$&$	-2.08 	\pm	0.02 	$&$	748 	\pm	24 	$&$	33.7 	\pm	1.9 	$&$	7.68 	\pm	0.36 	$&$	(	3.67 	\pm	0.60 	)\times	10^{25}	$&$	1.79 	$\\
NaI+BGO+LLE	&$[	187.20 	,	190.22 	]$&$	-0.69 	\pm	0.01 	$&$	-2.11 	\pm	0.02 	$&$	769 	\pm	23 	$&$	26.2 	\pm	1.6 	$&$	7.85 	\pm	0.36 	$&$	(	3.32 	\pm	0.54 	)\times	10^{25}	$&$	1.75 	$\\
\hline\hline																											$t_{\rm obs} ({\rm s})$ &	Time Interval (s)			&	$\alpha$			&	$\beta$			&	$E_0  ({\rm keV})$			&	$E_1 ({\rm MeV})$			&	$N_0$\tablenotemark{a}			&	$\Lambda ({\rm cm})$			&	$\chi_r^{2}$	\\
\hline
188.26 	&$[	187.20 	,	188.92 	]$&$	-0.69 	\pm	0.02 	$&$	-2.04 	\pm	0.03 	$&$	928 	\pm	46 	$&$	26.3 	\pm	2.1 	$&$	5.29 	\pm	0.38 	$&$	(	1.96 	\pm	0.51 	)\times	10^{25}	$&$	1.17 	$\\
188.75 	&$[	188.00 	,	189.50 	]$&$	-0.68 	\pm	0.01 	$&$	-2.18 	\pm	0.02 	$&$	855 	\pm	30 	$&$	33.6 	\pm	2.2 	$&$	9.19 	\pm	0.50 	$&$	(	9.74 	\pm	1.73 	)\times	10^{25}	$&$	1.46 	$\\
189.54 	&$[	188.92 	,	190.22 	]$&$	-0.70 	\pm	0.01 	$&$	-2.32 	\pm	0.03 	$&$	720 	\pm	25 	$&$	43.2 	\pm	4.4 	$&$	11.9 	\pm	0.7 	$&$	(	2.70 	\pm	0.69 	)\times	10^{26}	$&$	1.48 	$\\
\hline
\end{tabular}
\tablenotetext{a}{$N_0$ is in the unit of ${\rm photons\cdot cm^{-2}\cdot s^{-1}\cdot keV^{-1}}$.}
\end{table}
\section{Estimation of $\Lambda\equiv {R}/(2\Gamma)^{2\beta }$}\label{Appendix:B}
In this section, we present the derivation process of Equation~(\ref{Eq:R_Gamma}).
We first introduce two frames: the observer frame and
the comoving frame of the shell, denoted by a
prime, which is boosted radially with a Lorentz factor $\Gamma$ relative to the observer frame.
The radiation spectral power from per unit solid angle of the jet shell in the comoving frame of the shell is assumed as $P'_{E'}=dP'/dE'$.
Without considering the absorption of two-photon pair production,
the photon density in the jet shell comoving frame
can be described as
\begin{equation}\label{Eq:photon density}
dn'\simeq\frac{1}{4\pi R^2c}\frac{4\pi P'_{E'}}{E'}dE'
\equiv
A\left( {\frac{{E'}}{{{E'_{\rm ch}}}}} \right)^\beta dE',
\end{equation}
where $R$ is the radius of the shell from the central engine, $c$ is the light velocity,
and $4\pi R^2c \times 1\rm s$ is the filling volume for photons produced in one second.
Provided that the photons are emitted isotropically in the fluid frame,
the received power (without considering the absorption of two-photon pair production)
into a solid angle $\delta \Omega$
in the direction of the observer is given as
\begin{equation}\label{Eq:observed photons}
\delta P = \frac{1}{{d_L^2}}{D^3}(1+z)\frac{P'_{E'}dE}{{4\pi }}\delta \Omega,
\end{equation}
where $d_L$ is the luminosity distance and $D$ is the Doppler factor of the emitter.
With Equations~(\ref{Eq:photon density}) and (\ref{Eq:observed photons}),
the total observed photons (without considering the absorption of two-photon pair production)
at time $t_{\rm obs}$ can be described as
\begin{equation}
n_{\rm obs}(E,t_{\rm obs})dE = \int_{\rm (EATS)}{\frac{{{{(1 + z)}^2}A{R^2}c}}{{4\pi d_L^2}}} {D^2}{\left[ {\frac{{(1 + z)E}}{{D{{E'}_{{\rm{ch}}}}}}} \right]^\beta }d\Omega dE \simeq\frac{{{{(1 + z)}^2}A{R^2}c}}{{d_L^2}}{\left( {\frac{E}{{{E_{{\rm{ch}}}}}}} \right)^\beta }\frac{1}{{1 - \beta }}dE
\end{equation}
where EATS is the equal-arrival time surface
corresponding to the same observer time $t_{\rm obs}$
and $E_{\rm ch}=2\Gamma E'_{\rm ch}/(1 + z)$.
In Equation~(\ref{Eq:BandCutoff}), we describe the photon spectrum without considering
the absorption of two-photon pair production as
\begin{equation}
N_{E}= N_0K_2{\left( {\frac{E}{{{E_{\rm ch}}}}} \right)^\beta }\;{\rm and}\;E_{\rm ch}=1{\rm keV}.
\end{equation}
Then, one can have
\begin{equation}\label{Eq:A}
A = \frac{{{N_0}{K_2}(1 - \beta )d_L^2}}{{{(1 + z)^2R^2}c}}
\end{equation}
and
\begin{equation}\label{Eq:photon density_Final}
dn'\simeq\frac{{{N_0}{K_2}(1 - \beta )d_L^2}}{{{(1 + z)^2R^2}c}}\left( {\frac{{E'}}{{{E'_{\rm ch}}}}} \right)^\beta dE'.
\end{equation}
It should be noted that Equation~(\ref{Eq:photon density_Final}) is the same as the
equation~(3) in the Supporting Online Material of \cite{Abdo_AA-2009-Ackermann_M-Sci.323.1688A}.
For the equation~(3) in the Supporting Online Material of \cite{Abdo_AA-2009-Ackermann_M-Sci.323.1688A},
the factor of $(1+z)W'/(\Gamma c)$ with $W'$ being the jet shell width is the duration of observations.

The photoabsorption optical depth of high energy $\gamma$-rays ($\varepsilon'$)
from lower energy photons emitted cospatially in the jet shell is given by
(e.g., \citealp{Zhang_Yue-2019-Geng_JinJun-ApJ.877.89Z})
\begin{equation}
\tau _{\gamma \gamma }(\varepsilon')\simeq
\int {d\Omega' \int_{{\varepsilon'_c}}^\infty{\sigma _{\gamma\gamma}(\varepsilon',E',\theta')(1 - \cos\theta'){{\eta} W'}\frac{dn'}{4\pi}} }.
\end{equation}
Here, $\theta'$ is the incident angle,
$\varepsilon'_c$ is the lowest photon energy to perform pair production with $\varepsilon'$ photons,
i.e., ${\varepsilon'\varepsilon'_c(1 - \cos \theta ') = 2{{({m_{\rm{e}}}{c^2})}^2}}$,
$W'$ is the jet shell width in the comoving frame,
$\eta(\leqslant 1)$ describes the fraction of $W'$ making contribution to the pair production,
${{\sigma _{\gamma \gamma }}{\rm{(}}\varepsilon ',E',\theta ') = {\sigma _{\rm{T}}}g(y)}$ with
\begin{equation}
{y = \sqrt {1 - \frac{{2{{({m_{\rm{e}}}{c^2})}^2}}}{{\varepsilon 'E'(1 - \cos \theta ')}}} }
\end{equation}
and
\begin{equation}
{g(y) = \frac{3}{{16}}(1 - {y^2})\left[ {(3 - {y^4})\ln \frac{{1 + y}}{{1 - y}} - 2y(2 - {y^2})} \right]},
\end{equation}
and $m_e$ and $\sigma_{\rm T}$ are the  electron mass and Thomson cross section, respectively.\\
With $d{y^2} = 2(m_ec^2)^2dE'/[ \varepsilon'{E'}^2(1 - \cos \theta ')]$, one can have
\begin{equation}
\tau _{\gamma \gamma }(\varepsilon) =
W'A{\sigma _{\rm{T}}}{\left( {\frac{{1 + z}}{2\Gamma }} \right)^{ - 1 - 2\beta }}{E_{{\rm{ch}}}}{\left( {\frac{{{E_{{\rm{ch}}}}{\varepsilon}}}{{m_e^2{c^4}}}} \right)^{ - 1 - \beta }}F(\beta ),
\end{equation}
where
\begin{equation}
{F(\beta ) = \frac{4}{{1 - \beta }}\int_0^1 {{{\left( {1 - {y^2}} \right)}^{ - 2 - \beta }}yg(y)dy} }.
\end{equation}
By taking $W'=R/(2\Gamma)$ and setting $\tau _{\gamma \gamma }(E_c)=1$, one can have
\begin{equation}
\Lambda  \equiv
\frac{R}{{{{\left( {2\Gamma } \right)}^{2\beta }}}}
=\frac{{{N_0}{K_2}{\sigma _{\rm{T}}}d_L^2}}{2c}{E_{{\rm{ch}}}}(1 - \beta ){(1 + z)^{ - 2\beta  - 3}}{\left( {\frac{{{E_{{\rm{ch}}}}{E_c}}}{{m_e^2{c^4}}}} \right)^{ - 1 - \beta }}F(\beta ),
\end{equation}
where $\eta=1/2$ is adopted.
We would like to point out that \cite{Gupta_N-2008-Zhang_B-MNRAS.384L.11G}
was first suggested that the pair cutoff energy depends on both $\Gamma$ and $R$.

\section{$\Gamma$ estimated by adopting different division method on the first pulse in the second emission episode}\label{Sec:Different_Division}
\begin{figure}
\centering
\begin{tabular}{ccc}
\includegraphics[angle=0,scale=0.25,trim=100 0 50 0,clip]{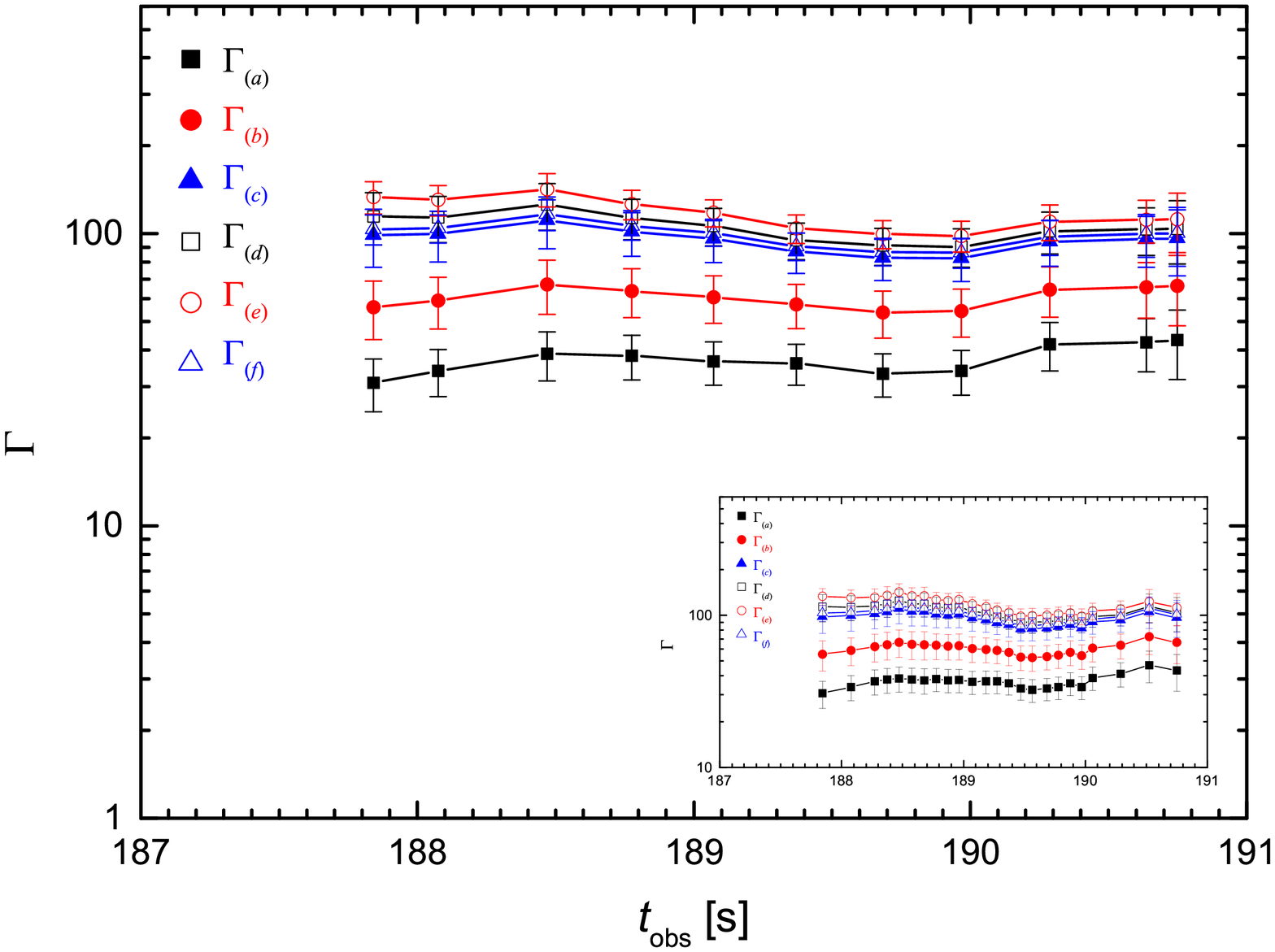} &
\includegraphics[angle=0,scale=0.25,trim=100 0 50 0,clip]{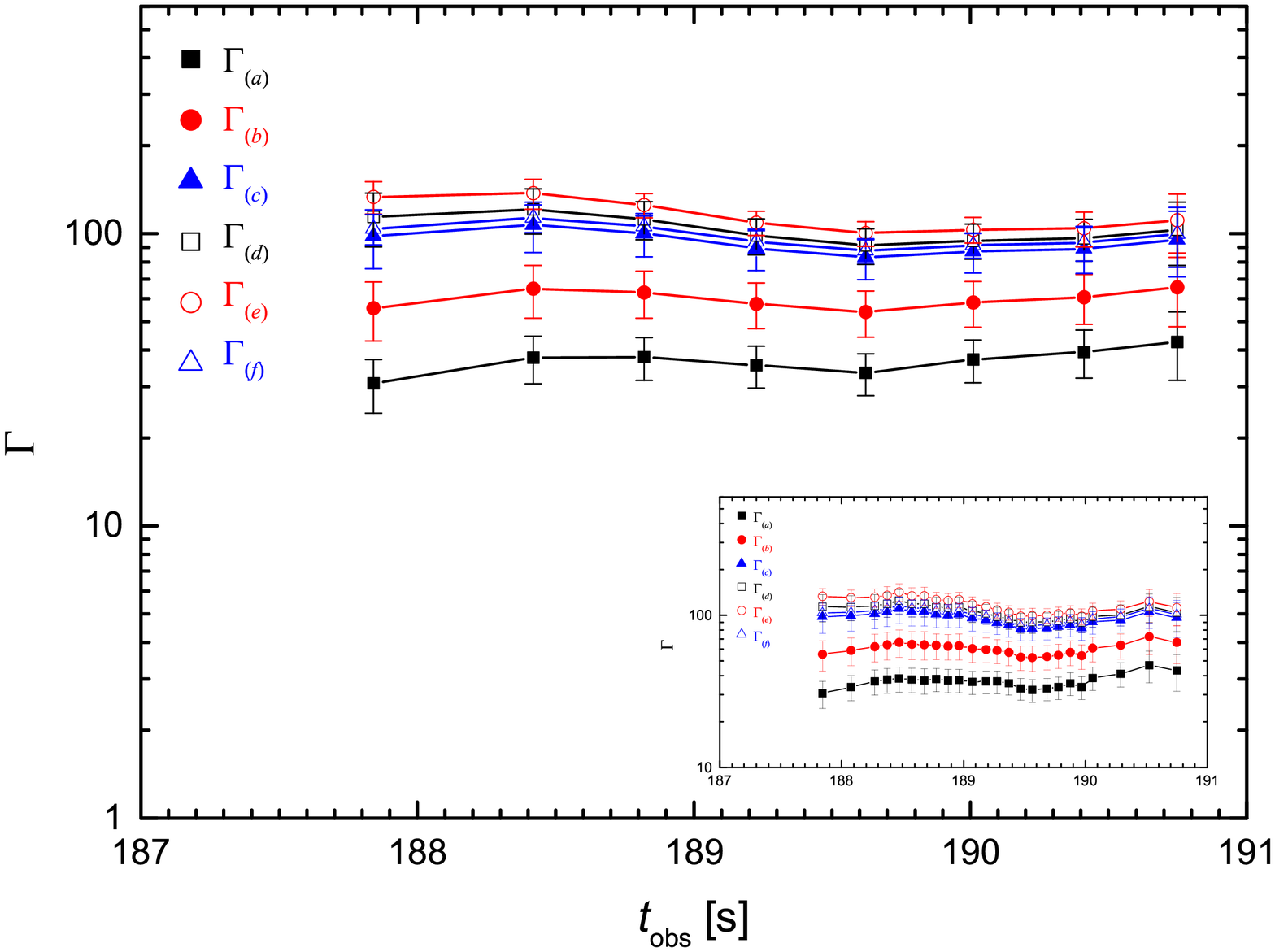} &
\includegraphics[angle=0,scale=0.25,trim=100 0 50 0,clip]{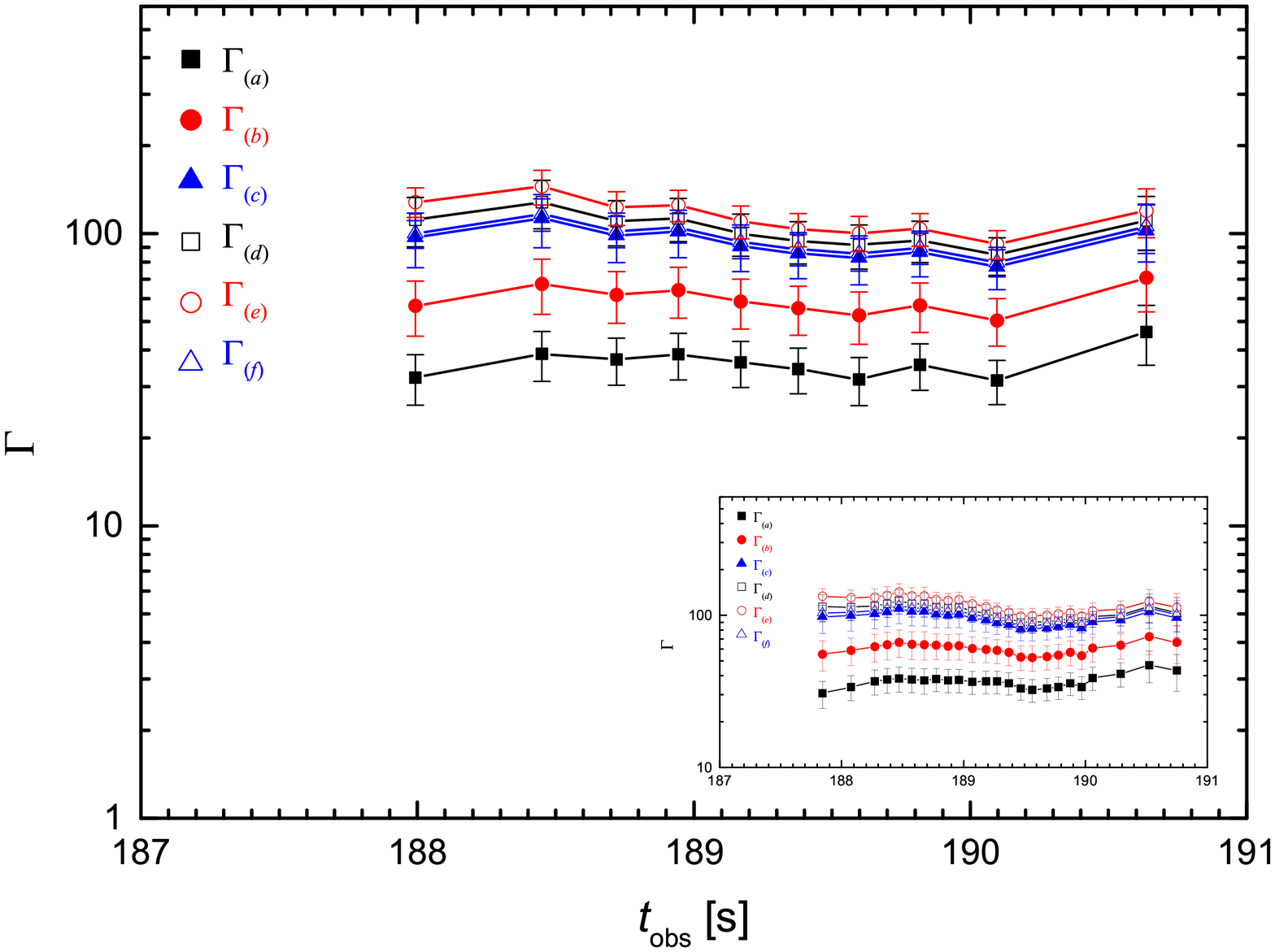} \\
\end{tabular}
\caption{Dependence of $\Gamma$ on $t_{\rm obs} (\neq 186.83\rm s)$,
where the different division methods, i.e., method I (left panel), II (middle panel), and III (right panel),
are adopting to divide the FP2EE of GRB~160625B.
The insets in each panel are the same as that in the left panel of Figure~\ref{Fig:R_Gamma}.}\label{Fig_App:DifferentDivision_GR}
\end{figure}

We also estimate the value of $\Gamma$ by adopting different division method on the FP2EE.
The time intervals, the joint spectral fitting results, and
the value of $\Lambda$ in each division method are shown in Table~\ref{Tab_App:DifferentDivision}.
It can be found that the results reported in Table~\ref{Tab_App:DifferentDivision} are consistent
with those in Tables~\ref{Tab:Fitting_Results}.
With Equation~(\ref{Eq:Differential_Relation}) and a given $\Gamma_{\rm try}=\Gamma(t_{\rm obs}=186.83\rm s)$,
we estimate the value of $\Gamma$ at different $t_{\rm obs}(\neq 186.83\rm s)$.
The results are shown in Figure~\ref{Fig_App:DifferentDivision_GR}.
One can find that the results shown in this figure are consistent with those in Figure~\ref{Fig:R_Gamma}.
Then, our obtained $\Gamma$ and its evolution with time are robust.

\clearpage
\begin{table}
{\centering
\caption{Spectral fitting results of the FP2EE in GRB~160625B and the value of $\Lambda$.}\label{Tab_App:DifferentDivision}
\begin{tabular}{c}
Division Method I\\
\end{tabular}
\begin{tabular}{c|cccccccc}
\hline\hline
$t_{\rm obs}(\rm s)$	&	Time Interval (s)			&	$\alpha$			&	$\beta$			&	$E_0  ({\rm keV})$			&	$E_1 ({\rm MeV})$			&	$N_0$\tablenotemark{a}			&	$\Lambda ({\rm cm})$			&	$\chi_r^{2}$	\\
\hline
186.83 	&$[	186.00 	,	187.50 	]$&$	-0.86 	\pm	0.06 	$&$	-1.56 	\pm	0.09 	$&$	1225 	\pm	393 	$&$	13.9 	\pm	0.0 	$&$	1.86 	\pm	0.49 	$&$	(	1.46 	\pm	1.03 	)\times	10^{23}	$&$	1.01 	$\\
187.84 	&$[	187.20 	,	188.25 	]$&$	-0.69 	\pm	0.03 	$&$	-1.83 	\pm	0.05 	$&$	873 	\pm	98 	$&$	20.3 	\pm	0.0 	$&$	2.78 	\pm	0.41 	$&$	(	2.39 	\pm	1.03 	)\times	10^{24}	$&$	1.09 	$\\
188.08 	&$[	187.71 	,	188.35 	]$&$	-0.69 	\pm	0.03 	$&$	-1.97 	\pm	0.06 	$&$	993 	\pm	96 	$&$	25.4 	\pm	0.0 	$&$	3.99 	\pm	0.54 	$&$	(	1.06 	\pm	0.49 	)\times	10^{25}	$&$	1.06 	$\\
188.47 	&$[	188.32 	,	188.62 	]$&$	-0.66 	\pm	0.03 	$&$	-1.99 	\pm	0.07 	$&$	890 	\pm	80 	$&$	22.9 	\pm	0.0 	$&$	7.78 	\pm	1.01 	$&$	(	2.20 	\pm	1.10 	)\times	10^{25}	$&$	1.08 	$\\
188.78 	&$[	188.62 	,	188.92 	]$&$	-0.69 	\pm	0.03 	$&$	-2.13 	\pm	0.07 	$&$	866 	\pm	66 	$&$	29.0 	\pm	0.0 	$&$	10.7 	\pm	1.2 	$&$	(	6.92 	\pm	3.52 	)\times	10^{25}	$&$	0.95 	$\\
189.07 	&$[	188.92 	,	189.22 	]$&$	-0.66 	\pm	0.03 	$&$	-2.14 	\pm	0.06 	$&$	730 	\pm	53 	$&$	26.3 	\pm	0.0 	$&$	10.9 	\pm	1.2 	$&$	(	6.25 	\pm	3.00 	)\times	10^{25}	$&$	1.06 	$\\
189.37 	&$[	189.22 	,	189.52 	]$&$	-0.71 	\pm	0.02 	$&$	-2.34 	\pm	0.07 	$&$	793 	\pm	53 	$&$	38.3 	\pm	0.0 	$&$	13.6 	\pm	1.4 	$&$	(	3.40 	\pm	2.10 	)\times	10^{26}	$&$	1.17 	$\\
189.68 	&$[	189.52 	,	189.82 	]$&$	-0.68 	\pm	0.03 	$&$	-2.24 	\pm	0.07 	$&$	614 	\pm	46 	$&$	30.3 	\pm	0.0 	$&$	11.7 	\pm	1.4 	$&$	(	9.81 	\pm	5.13 	)\times	10^{25}	$&$	1.02 	$\\
189.97 	&$[	189.82 	,	190.12 	]$&$	-0.75 	\pm	0.03 	$&$	-2.30 	\pm	0.07 	$&$	691 	\pm	58 	$&$	44.2 	\pm	0.0 	$&$	12.9 	\pm	1.6 	$&$	(	1.81 	\pm	1.14 	)\times	10^{26}	$&$	1.09 	$\\
190.29 	&$[	190.12 	,	190.48 	]$&$	-0.78 	\pm	0.03 	$&$	-2.51 	\pm	0.05 	$&$	686 	\pm	56 	$&$	139 	\pm	0 	$&$	12.6 	\pm	1.5 	$&$	(	2.69 	\pm	2.10 	)\times	10^{27}	$&$	1.06 	$\\
190.64 	&$[	190.42 	,	190.87 	]$&$	-0.74 	\pm	0.03 	$&$	-2.49 	\pm	0.05 	$&$	508 	\pm	41 	$&$	199 	\pm	106 	$&$	9.59 	\pm	1.28 	$&$	(	2.41 	\pm	2.21 	)\times	10^{27}	$&$	1.06 	$\\
190.75 	&$[	190.52 	,	191.00 	]$&$	-0.75 	\pm	0.03 	$&$	-2.52 	\pm	0.05 	$&$	501 	\pm	40 	$&$	226 	\pm	0 	$&$	9.18 	\pm	1.23 	$&$	(	3.38 	\pm	4.06 	)\times	10^{27}	$&$	1.10 	$\\
\hline
\end{tabular}\\
\begin{tabular}{c}
Division Method II\\
\end{tabular}
\begin{tabular}{c|cccccccc}
\hline\hline
$t_{\rm obs}(\rm s)$	&	Time Interval (s)			&	$\alpha$			&	$\beta$			&	$E_0  ({\rm keV})$			&	$E_1 ({\rm MeV})$			&	$N_0$\tablenotemark{a}			&	$\Lambda ({\rm cm})$			&	$\chi_r^{2}$	\\
\hline
186.83 	&$[	186.00 	,	187.50 	]$&$	-0.86 	\pm	0.06 	$&$	-1.56 	\pm	0.09 	$&$	1225 	\pm	393 	$&$	13.9 	\pm	0.0 	$&$	1.86 	\pm	0.49 	$&$	(	1.46 	\pm	1.03 	)\times	10^{23}	$&$	1.01 	$\\
187.84 	&$[	187.20 	,	188.25 	]$&$	-0.69 	\pm	0.03 	$&$	-1.83 	\pm	0.05 	$&$	873 	\pm	98 	$&$	20.3 	\pm	0.0 	$&$	2.78 	\pm	0.41 	$&$	(	2.39 	\pm	1.03 	)\times	10^{24}	$&$	1.09 	$\\
188.42 	&$[	188.22 	,	188.62 	]$&$	-0.69 	\pm	0.03 	$&$	-2.01 	\pm	0.06 	$&$	966 	\pm	78 	$&$	22.8 	\pm	3.3 	$&$	7.98 	\pm	0.90 	$&$	(	2.34 	\pm	1.08 	)\times	10^{25}	$&$	1.08 	$\\
188.82 	&$[	188.62 	,	189.02 	]$&$	-0.69 	\pm	0.02 	$&$	-2.13 	\pm	0.06 	$&$	855 	\pm	56 	$&$	28.2 	\pm	4.3 	$&$	11.2 	\pm	1.1 	$&$	(	6.57 	\pm	2.83 	)\times	10^{25}	$&$	1.02 	$\\
189.23 	&$[	189.02 	,	189.42 	]$&$	-0.66 	\pm	0.02 	$&$	-2.23 	\pm	0.06 	$&$	717 	\pm	43 	$&$	30.2 	\pm	5.4 	$&$	11.3 	\pm	1.1 	$&$	(	1.25 	\pm	0.58 	)\times	10^{26}	$&$	1.17 	$\\
189.62 	&$[	189.42 	,	189.82 	]$&$	-0.69 	\pm	0.02 	$&$	-2.26 	\pm	0.06 	$&$	665 	\pm	42 	$&$	31.4 	\pm	5.7 	$&$	12.5 	\pm	1.3 	$&$	(	1.18 	\pm	0.55 	)\times	10^{26}	$&$	1.12 	$\\
190.01 	&$[	189.82 	,	190.22 	]$&$	-0.75 	\pm	0.03 	$&$	-2.40 	\pm	0.05 	$&$	708 	\pm	50 	$&$	70.7 	\pm	18.2 	$&$	12.6 	\pm	1.4 	$&$	(	6.41 	\pm	3.61 	)\times	10^{26}	$&$	1.12 	$\\
190.41 	&$[	190.22 	,	190.62 	]$&$	-0.79 	\pm	0.03 	$&$	-2.50 	\pm	0.05 	$&$	638 	\pm	51 	$&$	132 	\pm	52 	$&$	12.2 	\pm	1.5 	$&$	(	1.95 	\pm	1.49 	)\times	10^{27}	$&$	1.00 	$\\
190.75 	&$[	190.52 	,	191.00 	]$&$	-0.75 	\pm	0.03 	$&$	-2.52 	\pm	0.05 	$&$	501 	\pm	40 	$&$	226 	\pm	162 	$&$	9.18 	\pm	1.23 	$&$	(	3.38 	\pm	4.06 	)\times	10^{27}	$&$	1.10 	$\\
\hline
\end{tabular}\\
\begin{tabular}{c}
Division Method III\\
\end{tabular}
\begin{tabular}{c|cccccccc}
\hline\hline
$t_{\rm obs}(\rm s)$	&	Time Interval (s)			&	$\alpha$			&	$\beta$			&	$E_0  ({\rm keV})$			&	$E_1 ({\rm MeV})$			&	$N_0$\tablenotemark{a}			&	$\Lambda ({\rm cm})$			&	$\chi_r^{2}$	\\
\hline
186.85 	&$[	186.00 	,	187.52 	]$&$	-0.85 	\pm	0.06 	$&$	-1.56 	\pm	0.09 	$&$	1142 	\pm	364 	$&$	13.8 	\pm	1.9 	$&$	1.86 	\pm	0.49 	$&$	(	1.40 	\pm	0.96 	)\times	10^{23}	$&$	1.02 	$\\
187.99 	&$[	187.52 	,	188.31 	]$&$	-0.69 	\pm	0.03 	$&$	-1.92 	\pm	0.06 	$&$	942 	\pm	96 	$&$	23.5 	\pm	2.9 	$&$	3.35 	\pm	0.47 	$&$	(	5.93 	\pm	2.67 	)\times	10^{24}	$&$	1.01 	$\\
188.45 	&$[	188.31 	,	188.60 	]$&$	-0.66 	\pm	0.03 	$&$	-1.96 	\pm	0.07 	$&$	888 	\pm	83 	$&$	21.3 	\pm	3.3 	$&$	7.55 	\pm	1.01 	$&$	(	1.71 	\pm	0.86 	)\times	10^{25}	$&$	1.04 	$\\
188.72 	&$[	188.60 	,	188.84 	]$&$	-0.68 	\pm	0.03 	$&$	-2.13 	\pm	0.08 	$&$	878 	\pm	76 	$&$	25.6 	\pm	5.1 	$&$	9.90 	\pm	1.28 	$&$	(	5.97 	\pm	3.52 	)\times	10^{25}	$&$	0.95 	$\\
188.94 	&$[	188.84 	,	189.06 	]$&$	-0.70 	\pm	0.03 	$&$	-2.15 	\pm	0.07 	$&$	835 	\pm	73 	$&$	32.8 	\pm	7.0 	$&$	12.7 	\pm	1.6 	$&$	(	8.77 	\pm	4.98 	)\times	10^{25}	$&$	0.92 	$\\
189.17 	&$[	189.06 	,	189.27 	]$&$	-0.64 	\pm	0.03 	$&$	-2.25 	\pm	0.08 	$&$	694 	\pm	55 	$&$	34.2 	\pm	8.8 	$&$	10.2 	\pm	1.4 	$&$	(	1.64 	\pm	1.04 	)\times	10^{26}	$&$	0.97 	$\\
189.38 	&$[	189.27 	,	189.49 	]$&$	-0.70 	\pm	0.03 	$&$	-2.26 	\pm	0.09 	$&$	767 	\pm	62 	$&$	29.0 	\pm	7.0 	$&$	13.4 	\pm	1.7 	$&$	(	1.42 	\pm	0.93 	)\times	10^{26}	$&$	1.11 	$\\
189.60 	&$[	189.49 	,	189.71 	]$&$	-0.67 	\pm	0.03 	$&$	-2.15 	\pm	0.08 	$&$	623 	\pm	56 	$&$	19.8 	\pm	3.9 	$&$	11.6 	\pm	1.6 	$&$	(	3.69 	\pm	2.13 	)\times	10^{25}	$&$	0.96 	$\\
189.82 	&$[	189.71 	,	189.92 	]$&$	-0.76 	\pm	0.03 	$&$	-2.30 	\pm	0.08 	$&$	734 	\pm	69 	$&$	45.6 	\pm	13.5 	$&$	15.6 	\pm	2.1 	$&$	(	2.17 	\pm	1.50 	)\times	10^{26}	$&$	1.05 	$\\
190.10 	&$[	189.92 	,	190.32 	]$&$	-0.71 	\pm	0.03 	$&$	-2.27 	\pm	0.07 	$&$	606 	\pm	47 	$&$	38.0 	\pm	9.0 	$&$	9.96 	\pm	1.22 	$&$	(	1.02 	\pm	0.57 	)\times	10^{26}	$&$	1.10 	$\\
190.64 	&$[	190.32 	,	191.02 	]$&$	-0.77 	\pm	0.03 	$&$	-2.53 	\pm	0.04 	$&$	545 	\pm	35 	$&$	273 	\pm	166 	$&$	10.4 	\pm	1.1 	$&$	(	5.55 	\pm	5.64 	)\times	10^{27}	$&$	1.02 	$\\
\hline																										
\end{tabular}\\}
\tablenotemark{a}{$N_0$ is in the unit of ${\rm photons\cdot cm^{-2}\cdot s^{-1}\cdot keV^{-1}}$.}
\end{table}

\clearpage


\end{document}